\documentclass[a4paper,11pt]{article}

\usepackage{jcappub}

\usepackage[english]{babel}

\usepackage{booktabs}

\def\mean#1{\left\langle#1\right\rangle}

\title{Reconstruction of inclined air showers detected with the
Pierre Auger Observatory}

\collaboration{The Pierre Auger Collaboration\\}

\author[42]{A.~Aab} 
\author[65]{P.~Abreu} 
\author[54]{M.~Aglietta}
\author[95]{M.~Ahlers} 
\author[83]{E.J.~Ahn}
\author[29]{I.~Al Samarai} 
\author[17]{I.F.M.~Albuquerque}
\author[1]{I.~Allekotte} 
\author[87]{J.~Allen} 
\author[89]{P.~Allison} 

% ...
\affiliation[1]{ Centro At\'{o}mico Bariloche and Instituto Balseiro (CNEA-UNCuyo-CONICET), San 
Carlos de Bariloche, 
Argentina }
% ...
\affiliation[17]{ Universidade de S\~{a}o Paulo, Instituto de F\'{\i}sica, S\~{a}o Paulo, SP, 
Brazil }
% ...
\affiliation[29]{ Institut de Physique Nucl\'{e}aire d'Orsay (IPNO), Universit\'{e} Paris 11, CNRS-IN2P3, 
Orsay, 
France }
% ...
\affiliation[42]{ Universit\"{a}t Siegen, Siegen, 
Germany }
% ...
\affiliation[54]{ Osservatorio Astrofisico di Torino  (INAF), Universit\`{a} di Torino and Sezione INFN, 
Torino, 
Italy }
% ...
\affiliation[65]{ Laborat\'{o}rio de Instrumenta\c{c}\~{a}o e F\'{\i}sica Experimental de Part\'{\i}culas - LIP and  
Instituto Superior T\'{e}cnico - IST, Universidade de Lisboa - UL, 
Portugal }
% ...
\affiliation[83]{ Fermilab, Batavia, IL, 
USA }
% ...
\affiliation[87]{ New York University, New York, NY, 
USA }
% ...
\affiliation[89]{ Ohio State University, Columbus, OH, 
USA }
% ...
\affiliation[95]{ University of Wisconsin, Madison, WI, 
USA }
% ...
\affiliation[(c)] { Now at NYU Abu Dhabi }
% ...

\emailAdd{auger\_spokepersons@fnal.gov}

\abstract{We describe the method devised to reconstruct inclined cosmic-ray
  air showers with zenith angles greater than $60^\circ$ detected with the
  surface array of the Pierre Auger Observatory. The measured signals at the
  ground level are fitted to muon density distributions predicted with
  atmospheric cascade models to obtain the relative shower size as an overall
  normalization parameter. The method is evaluated using simulated showers to
  test its performance. The energy of the cosmic rays is calibrated using a
  sub-sample of events reconstructed with both the fluorescence and surface
  array techniques. The reconstruction method described here provides the
  basis of complementary analyses including an independent measurement of the
  energy spectrum of ultra-high energy cosmic rays using very inclined events
  collected by the Pierre Auger Observatory.}


\begin{document}
\maketitle
\flushbottom

\section{Introduction}
\normalsize 

The Pierre Auger Observatory is a hybrid instrument combining an array of
particle detectors, the Surface Detector (SD) array, to sample the air shower
front as it reaches the ground and Fluorescence Detector (FD) telescopes to
capture the ultraviolet light emitted by the nitrogen as showers develop in
the atmosphere. The FD is used to monitor the atmosphere, on dark clear
nights, above the 3000\,km$^2$ area over which the SD is laid out. The site is
located near the town of Malarg\"ue, in the Argentinian province of Mendoza,
at an altitude of about 1400\,m above sea level and at an average latitude of
$35.2^\circ$\,S~\cite{performance}.
 
The SD stations are water-Cherenkov detectors which are sensitive to inclined
particles, so that the array is also sensitive to very inclined showers. Since
the beginning of its deployment the SD of the Pierre Auger Observatory has
been routinely recording events with zenith angles up to $90^\circ$. However,
the reconstruction of events with zenith angles exceeding ${\sim}60^\circ$
requires a different method to the one used for events nearer the vertical,
due to an asymmetry induced in the lateral distribution of the shower
particles by the geomagnetic field.

The showers produced by cosmic-ray hadrons at large zenith angles traverse
much larger atmospheric depths than vertical showers and, as a result, their
shower maximum occurs higher in the atmosphere. By the time the shower front
reaches the ground, the generation of electrons and photons in the main
cascading process is basically finished, and the bulk of them have been
absorbed in the atmosphere. Most of the particles that reach the ground are
energetic muons from charged pion decays in the showering process accompanied
by a smaller electron and photon component that stems from the muons
themselves, primarily from decays in flight. These muons travel long distances
deviating in the geomagnetic field, so that when they reach the ground the
characteristic cylindrical symmetry of the showers is lost.

Although the interest in inclined showers dates to the early days of extensive
air shower measurements~\cite{Hillas}, it was only in the beginning of the
2000s, when the patterns of the muons at the ground level were sufficiently
understood~\cite{model}, that methods were developed to reconstruct data in a
reliable way~\cite{Ave}. The analysis of these showers is of particular
interest because it enhances the exposure of the detector by 30\% and extends
the sky coverage to regions that are otherwise unobservable. It provides the
basis of a completely independent measurement of the cosmic ray spectrum above
$4{\times}10^{18}$\,eV~\cite{VMOGThesis,DembinskiThesis,RAVicrc,HDicrc,AlexICRC},
to be updated in a forthcoming publication. Since inclined showers are mainly
composed of muons, their study provides an almost uncontaminated measurement
of the muon content of the shower, while for events with zenith angles less
than about $60^\circ$ the muonic component can not be disentangled from the
electromagnetic activity, and thus must be inferred indirectly in the absence
of shielded detectors, which is intrinsically more challenging.  Therefore,
the inclined data also provide complementary information to constrain the
nature of the arriving particles~\cite{PRL,AveAuger}.  In addition, they also
constitute the background against which the search for high-energy neutrinos
with inclined showers must be made~\cite{Berezinskii,Capelle}.

This article deals with inclined showers as follows. Section~\ref{s:sd}
describes details of the Surface Detector of the Pierre Auger Observatory that
are of relevance for inclined-shower reconstruction. Section~\ref{s:modeling}
deals with the modeling needed for the reconstruction procedure, namely the
muon distribution at the ground level, the treatment of the electromagnetic
contribution to the signal and also the signal response of the surface
detectors to the passage of muons. In section~\ref{s:reconstruction} the
details of the reconstruction procedure are described and the uncertainties
addressed. The energy calibration procedure is discussed in
section~\ref{s:energy_calibration}, and section~\ref{s:conclusions} summarizes
the conclusions.

\section{Inclined events in the Auger Observatory}
\label{s:sd}

\begin{figure}[tbp]
\centering
\includegraphics[width=0.47\textwidth]{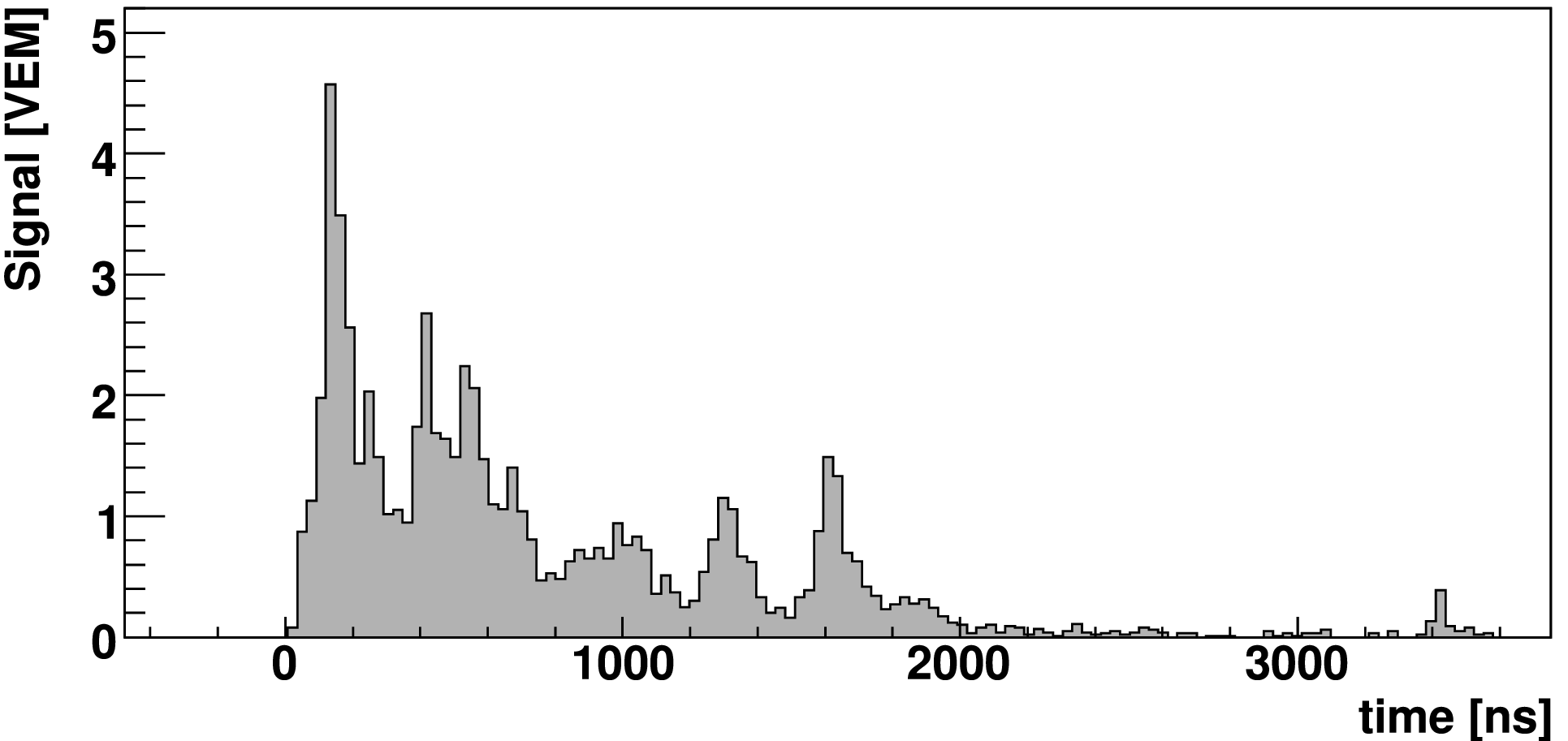}
\hfill
\includegraphics[width=0.47\textwidth]{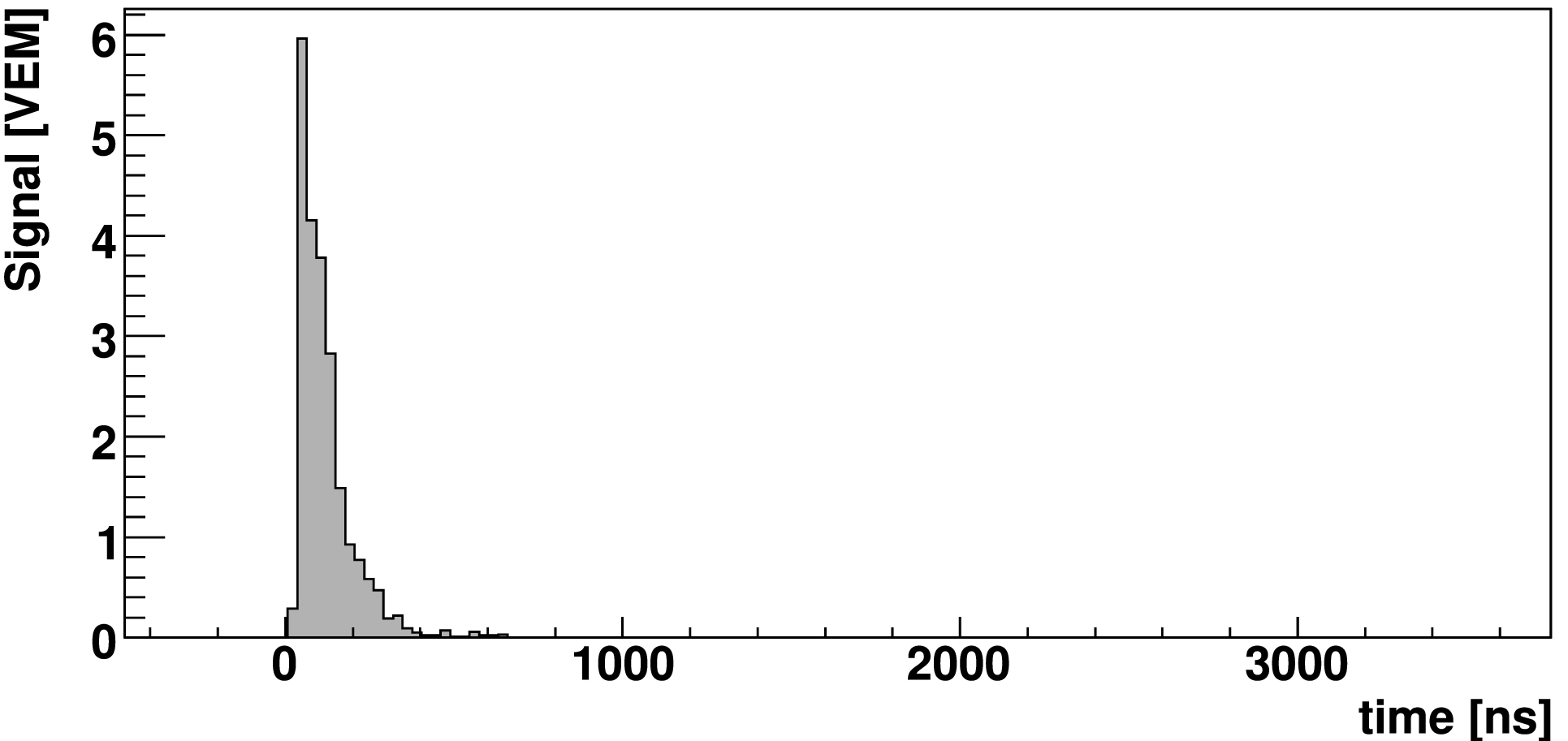}
\vspace{0.3cm}
\caption{\label{FADCs} FADC traces of SD stations at 1 km from the shower core
  in units of VEM. \emph{Left:} signal characteristic of a nearly vertical
  shower (zenith angle ${\sim}22^\circ$), where the electromagnetic component
  contributes about ${\sim}50$\% to the total signal size. \emph{Right:}
  signal characteristic of an inclined shower (zenith angle ${\sim}80^\circ$),
  where the muonic component dominates the total signal.}
\end{figure}

The SD consists of more than 1600 water-Cherenkov stations arranged on a
hexagonal grid, each station at a distance of 1.5\,km from its six nearest
neighbors.  Each station is a cylindrical water tank of 10\,m$^2$ surface
area, filled to 1.2\,m with purified water contained inside a
diffusely-reflective liner. The volume of water is viewed by three 9\,inch
photo-multipliers that detect the Cherenkov light emitted during the passage
of charged particles. The signal is digitized in time slots of 25\,ns using a
Flash Analog-to-Digital Converter (FADC) running at 40\,MHz. All the stations
are controlled remotely, and data are transmitted from the detectors to a
central station by Local Area Network radio links. Synchronization to 8\,ns of
relative precision is provided by commercial GPS systems. A full description
of the SD can be found in~\cite{performance}. The stations are calibrated
on-line continuously by identifying the maximum in the raw signal histograms
of signals produced by atmospheric background muons which are sampled at
regular intervals.  The integrated charge signal of this maximum is related to
that of a vertical muon traversing the detector through its center,
\emph{Vertical Equivalent Muon} or VEM, which provides the signal unit with
3\% accuracy~\cite{calibration}. The total signal at each station in VEM
units, $S^\text{meas}$, is obtained by integrating the signal traces in
time. Examples of FADC traces in VEM units are shown in figure~\ref{FADCs}.

When shower electrons and photons ($\mean{E}\ge10$\,MeV) reach the surface
detectors they are absorbed in the water and provide a Cherenkov light signal
which is approximately proportional to the total deposited energy. Shower
muons are more energetic ($E\ge250$\,MeV) and travel right through the tanks,
typically giving signals proportional to their track length. For vertical
showers initiated by protons or nuclei, the contribution of electrons and
photons to the signal is often comparable or even larger than that due to the
muons~\cite{Ines}. However, for inclined showers muons dominate the signal due
to the electromagnetic component being largely absorbed by the
atmosphere. Moreover the detector increases its relative response to muons
with respect to electrons and photons as the zenith angle increases since the
muon signal scales with the track length and the diameter of the tank exceeds
its height.

The FD is distributed in four buildings on the perimeter of the surface array.
Each building contains six telescopes that together cover $180^\circ$ in
azimuth and nearly $30^\circ$ in elevation. The FD is fully described
in~\cite{FD}.  Whereas the SD measurements are performed with a duty cycle of
almost 100\%, the FD only operates on clear moonless nights and has a duty
cycle of 13\%. The FD allows a calorimetric measurement of the shower energy
deposited in the atmosphere, in contrast to the SD. The positions of the
triggered pixels and the arrival time of the light are used to extract the
shower direction. The profile function of the energy deposited in the
atmosphere is determined from the signals at the triggered
pixels~\cite{FDRecons} after taking into account the separate fluorescence and
Cherenkov light contributions, as well as light attenuation and dispersion in
the atmosphere. The atmospheric conditions are monitored regularly using
several techniques to provide relevant data on attenuation and light
dispersion parameters~\cite{monitoring}. The electromagnetic energy released
by the shower in the atmosphere is obtained by fitting the longitudinal
profile to a Gaisser-Hillas function~\cite{GH} and integrating over the range of
atmospheric depths. The total energy of the primary particle is derived from
the calorimetric energy by adding the invisible energy which accounts for the
energy carried by penetrating particles. Most of the detected FD events also
induce triggers of at least one SD station and are called hybrid events. The
geometry of the shower axis can be determined more precisely for these events
by using timing information from the FD pixels, coupled with the arrival time
of the shower at the SD station with highest signal. A fraction of these
hybrid events, called ``golden hybrid'' events, have sufficient SD stations
triggered to allow for an independent reconstruction with SD techniques.

\begin{figure}[tbp]
\centering
\includegraphics[width=0.45\textwidth]{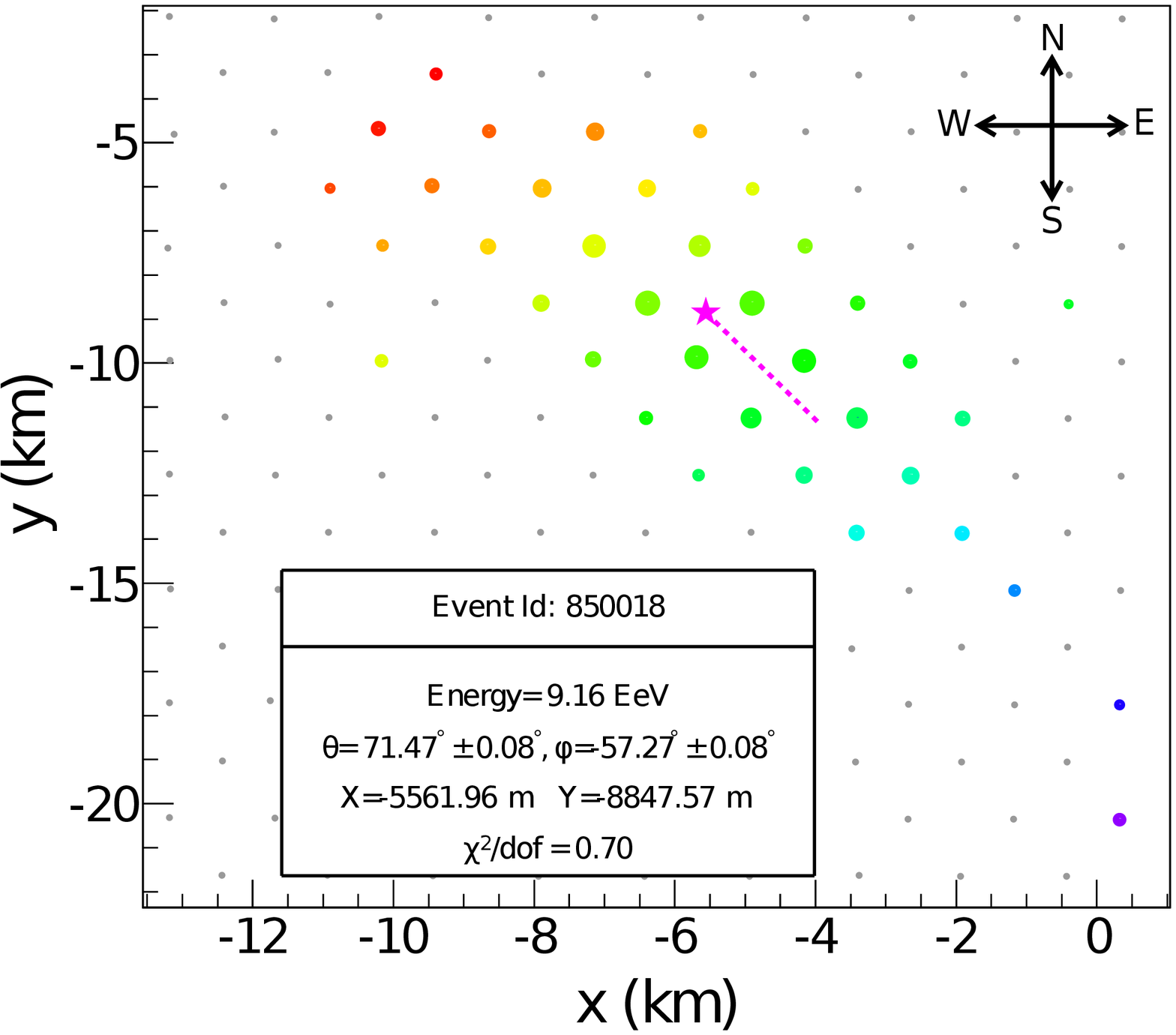}
\hfill
\includegraphics[width=0.45\textwidth]{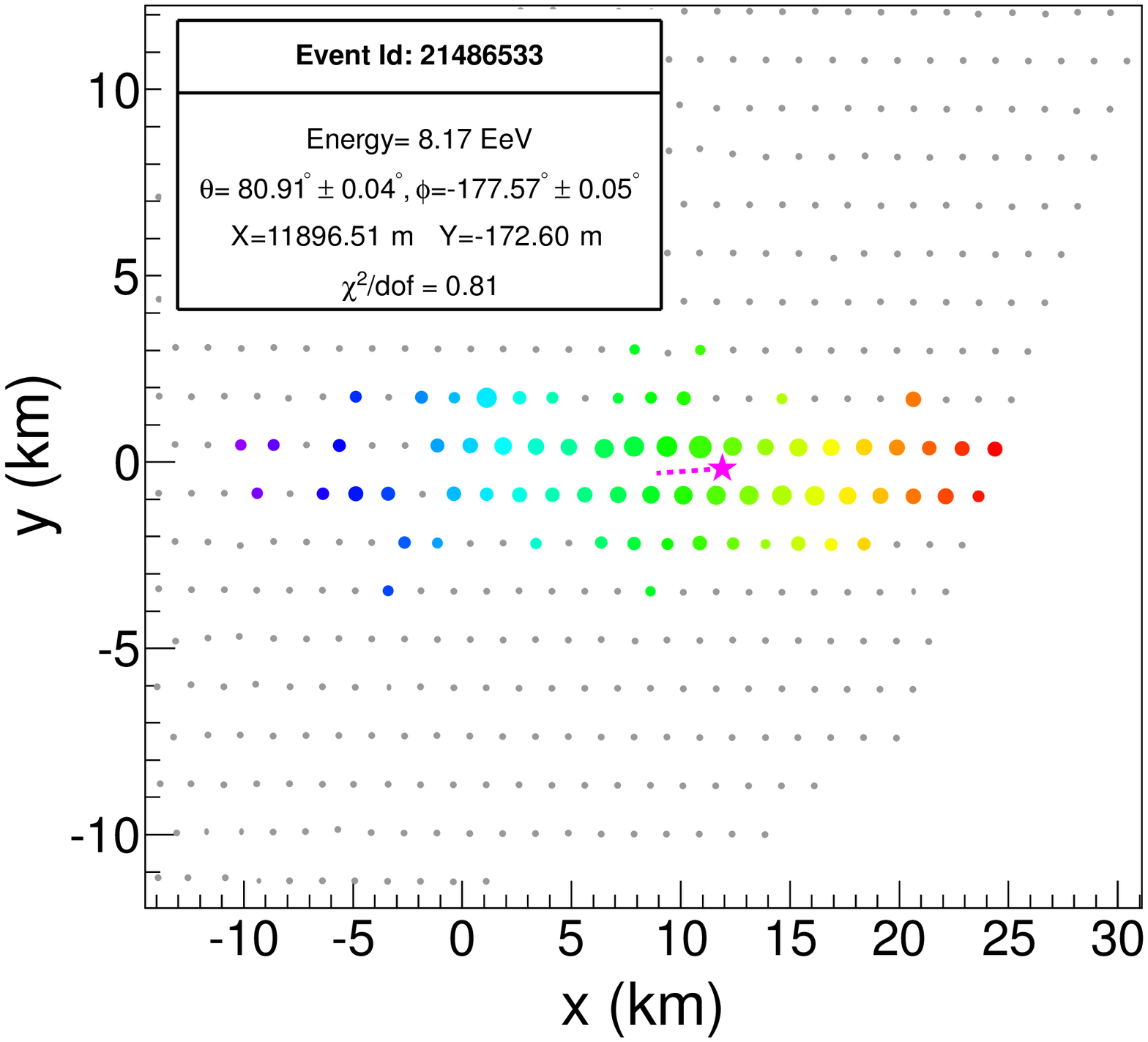}
\caption{Inclined events with large number of triggered stations. The left
  (right) event with 37 (69) triggered stations corresponds to zenith angle
  $71^\circ$ ($81^\circ$) and azimuth angle to the East $-57^\circ$
  ($-178^\circ$). The dashed line shows the shower axis projected onto the
  ground and the star indicates the core position. The areas of the circles
  are proportional to the logarithm of the signal size and the color code
  denotes the measured arrival time from early (blue) to late (red) stations.}
\label{LargeEvents}
\end{figure} 

\subsection{Selection of inclined events with the Surface Detector} 
 
The SD trigger (described in detail in~\cite{Piera}) has been designed in a
hierarchical pattern to identify a cosmic-ray event and reject random
coincidences. The first two triggers, T1 and T2, apply at station level. T1
and T2 require that the pulse height of the signal exceeds two preset values
(larger for T2 than for T1). Alternatively a second T2 trigger can also be
obtained with smaller pulses that are spread in time (T2-ToT), which is a
characteristic of the electromagnetic component in showers of near vertical
incidence (see left panel of figure~\ref{FADCs}). The T1 trigger is stored
temporarily while the data of all T2 triggers are sent to the central
station. T2 triggered stations that lie in a close and compact configuration
are checked for coincidences in time. If sufficient coincident stations are
found, the third level trigger (T3) is passed and the acquisition process
starts. In this process the T1 triggers stored at the local stations are also
downloaded and added to the event.

Starting with the T3 triggers, a higher-level trigger hierarchy is implemented
to satisfy the precision requirements of the consequent analysis.  The T4 and
T5 filters are respectively applied off-line for selecting physical events,
and to ensure that they fall on a region of the array where the surface
detectors were operational at the time of the event, to guarantee their
quality.  They are defined differently for inclined events. The ``inclined
T4'' condition ensures that the start time of the signals\footnote{Start time
  is defined as the arrival time of the first shower particle into the SD
  station.} of at least four nearby T2 triggered stations are compatible with
a shower front moving at the speed of light. Accidental triggers are removed
by eliminating stations that have times outside the window corresponding to
the passage of the shower front, as determined by the rest of the
stations. Stations are eliminated one by one (selecting the ones with the
highest timing offsets) starting from the T3 trigger selection, until a
satisfactory configuration with four or more stations in a compact arrangement
is found. This removes a large number of showers reconstructed with incorrect
arrival directions due to random coincidences. Events that have stations which
are all aligned (according to the hexagonal pattern of the array) are excluded
since the arrival direction cannot be reconstructed accurately from these
start-time data.

The inclined T5 condition is defined so as to select events that fall inside a
region of the array in which all stations were operational (``active''),
requiring that the station that is nearest to the reconstructed core and its
six adjacent stations are all operational. This active region defines an
``active unit cell'' shaped as a hexagon and forms the basis of the array
aperture calculation~\cite{Piera}. The T5 condition serves two purposes: (i)
it avoids the reconstruction of events that fall near to the edge of the array
or in regions where a station is temporarily not fully operational, which can
have large uncertainties, and (ii) it ensures that the shower falls inside the
active area that has been considered to establish the aperture. Note that the
T5 condition is the basic trigger of the selection criteria used for the
subsequent analysis of inclined showers.

\begin{figure}[tbp]
\centering
\includegraphics[height=0.3\textwidth]{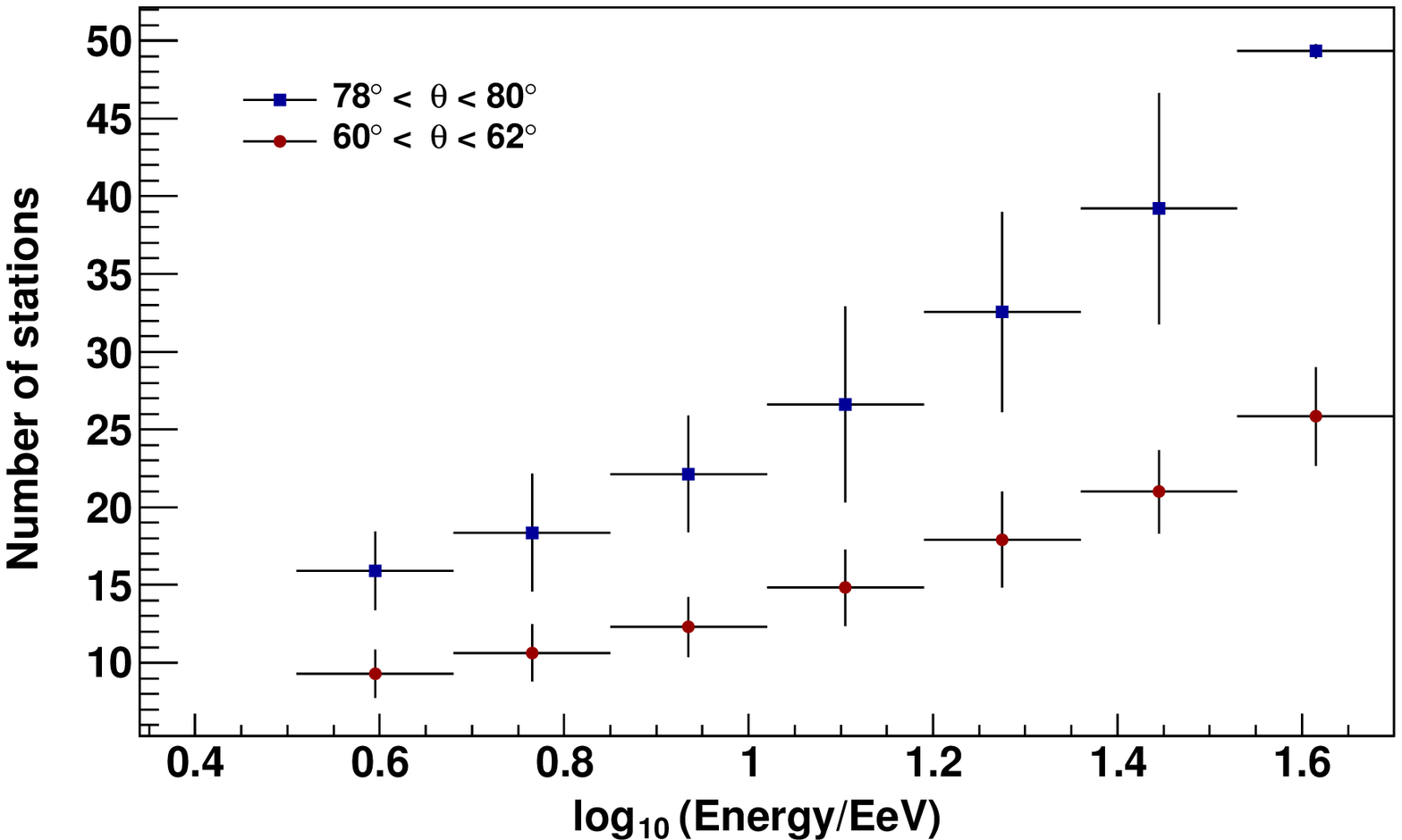}
\hfill
\includegraphics[height=0.305\textwidth]{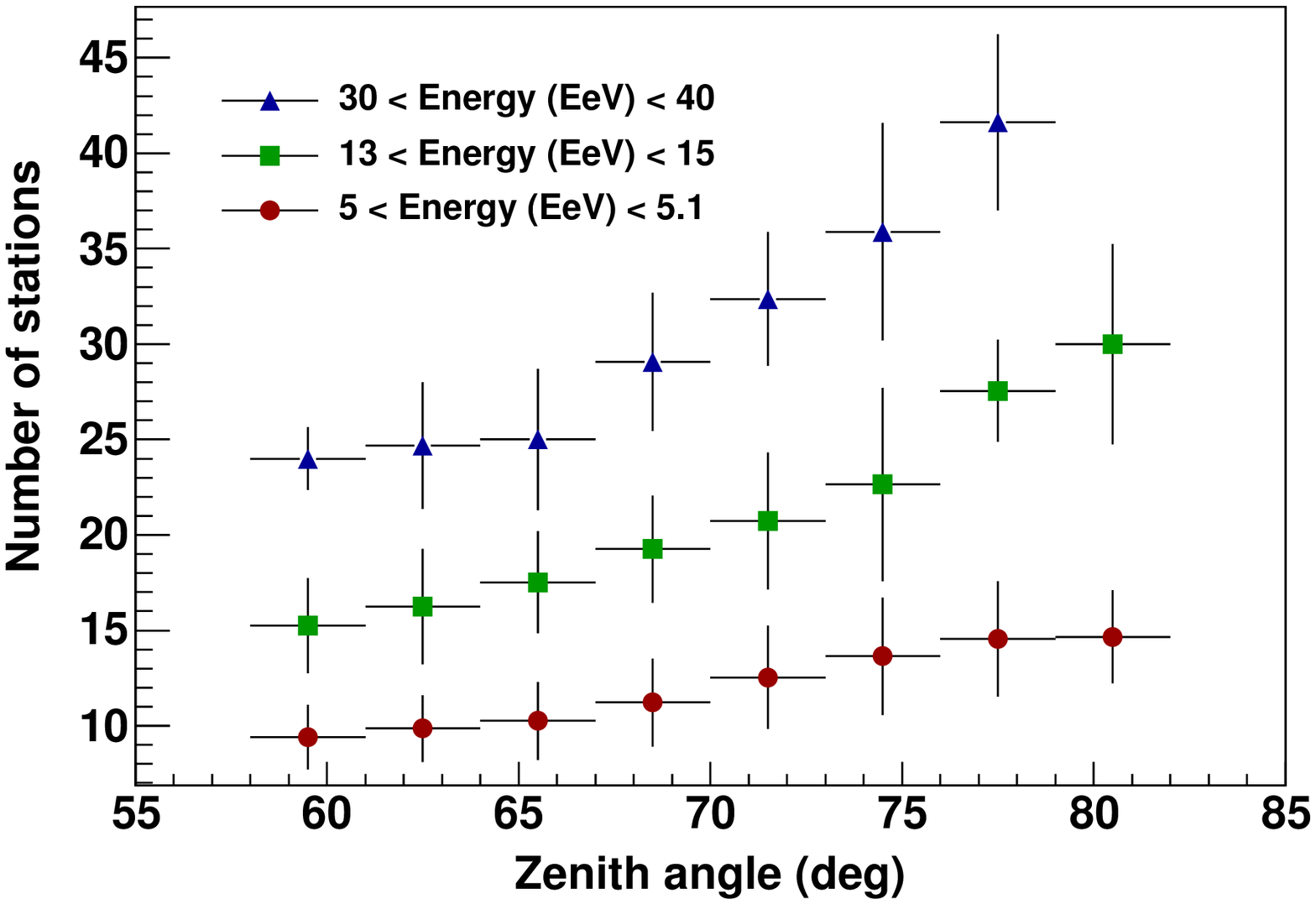}
\caption{\emph{Left:} average number of triggered stations as a function of
  shower energy. \emph{Right:} the same as a function of the zenith angle. The
  vertical error bars correspond to the standard deviation of the number of
  stations.}
\label{NTankAverages}
\end{figure} 

\subsection{General characteristics of inclined SD events} 

As the zenith angle $\theta$ of the showers increases, and they traverse
larger atmospheric depths, the showers that reach the ground are increasingly
attenuated. It appears that this would result in a reduced efficiency to
detect inclined showers with the SD with increasing zenith angle. On the
contrary, there is a compensating geometric effect since the density of
stations in the \emph{shower plane} (perpendicular to the shower axis)
increases rapidly in proportion to $\sec\theta$. The signal reduction due to
the muon attenuation over long path lengths is relatively small and is
compensated by some of the surface detectors being closer to the shower axis
where the signal is higher (compared to more vertical events) and thus more
likely to trigger. Inclined events can have a large number of triggered
stations as illustrated in figure~\ref{LargeEvents}.

This effect manifests itself when considering the number of triggered stations
of events with a given energy as a function of zenith angle. For instance,
considering an energy around 10\,EeV (the energy at which the SD is fully
efficient for events out to $80^\circ$), a vertical event has on average 8
triggered stations. This number is around 13 for a $60^\circ$ event of the
same energy, while at $80^\circ$ it increases to 25.  This effect is
illustrated in figure~\ref{NTankAverages} which shows the average number of
stations as a function of shower energy at a fixed zenith angle (left panel)
and the average number of stations as a function of zenith angle at a fixed
energy (right panel).

\section{Modeling particle distributions and detector responses}
\label{s:modeling}

Showers with $\theta>60^\circ$ require specific reconstruction methods since
they are dominated by muons displaying complex density patterns at the ground
level. The method used for the reconstruction of inclined showers is based on
a fit of the measured signals to the expected pattern, and requires modeling
of a two-dimensional distribution of the muon number densities at the ground,
the response of the detectors to the passage of muons, and the treatment of
the electromagnetic component of the signal in the detectors.

\begin{figure}[tbp]
\centering
\includegraphics[width=0.49\textwidth]{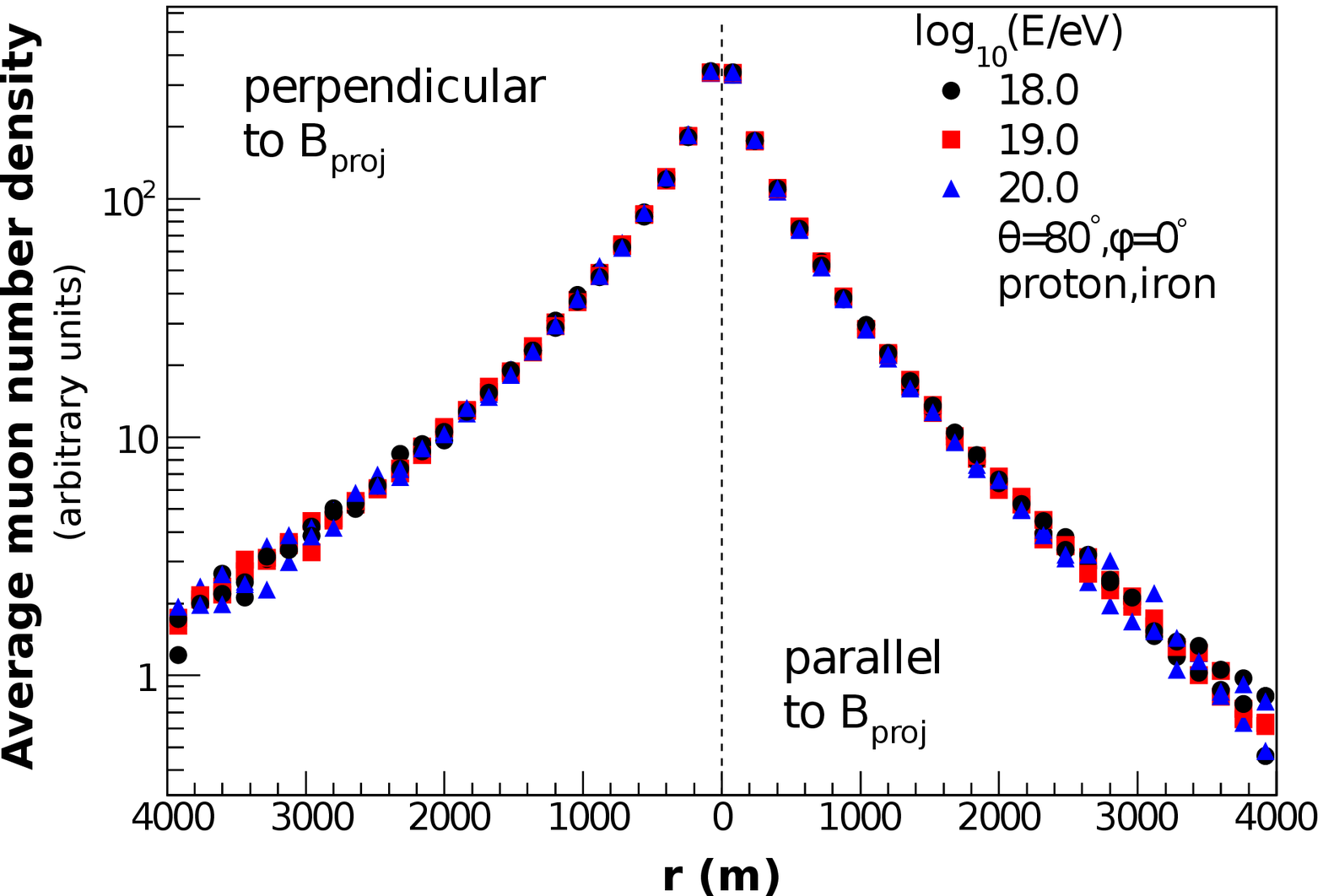}
\hfill
\includegraphics[width=0.47\textwidth]{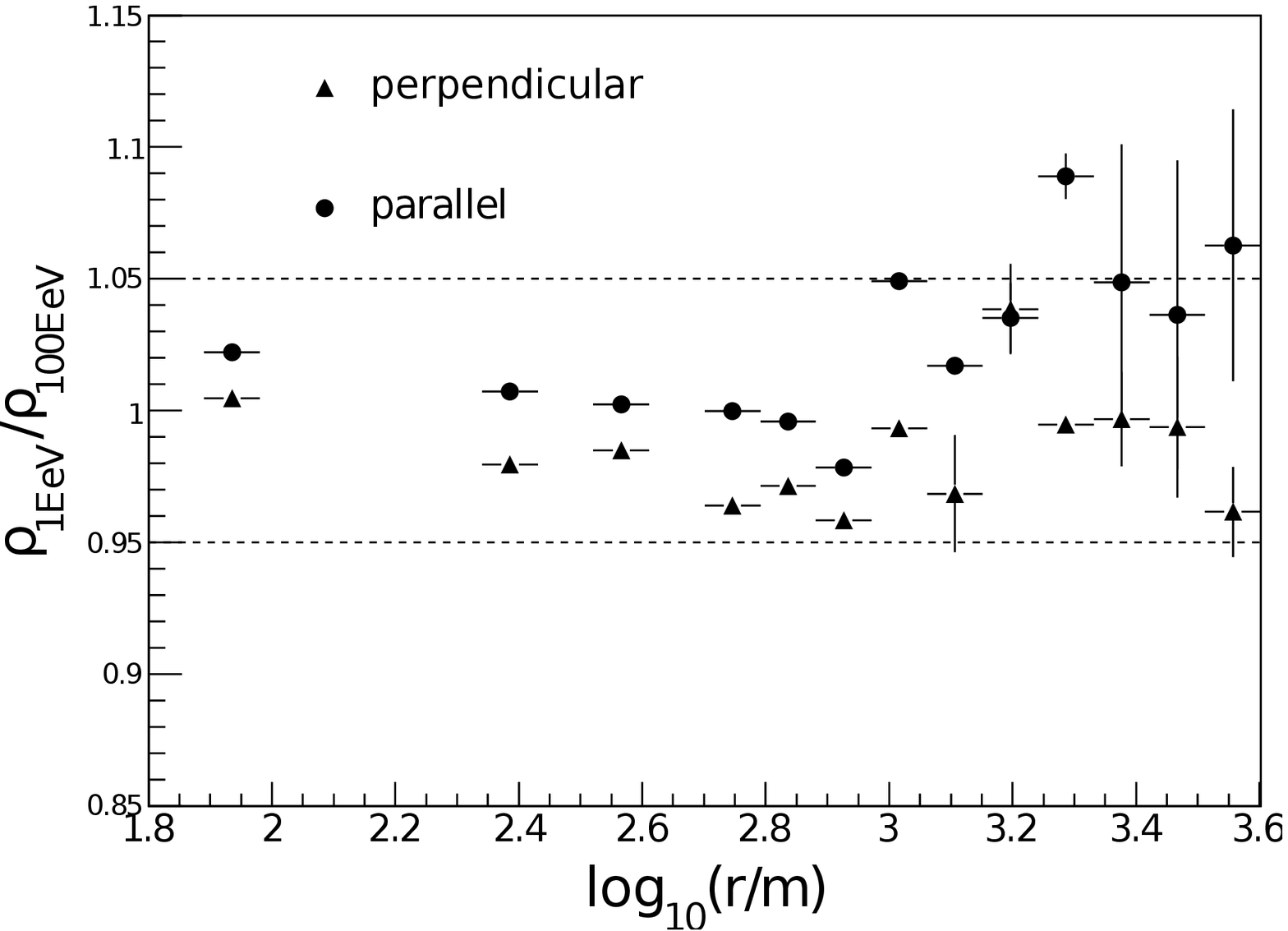}
\caption{\emph{Left:} average muon number density as a function of distance to
  the shower axis ($r$) in the shower plane, for showers with zenith angle
  $80^\circ$ and azimuth angle $0^\circ$ pointing to the East. The right
  (left) part of the $r$-axis corresponds to the direction parallel
  (perpendicular) to the projection of the geomagnetic field $\vec{B}$ onto
  the shower plane, $B_\text{proj}$. The distributions have been normalized to
  the equivalent density of $10^{19}$\,eV to illustrate the degree to which
  the shapes are compatible, for three different energies and two primary
  compositions, as labeled. \emph{Right:} ratio of average muon densities for
  1 and 100\,EeV proton showers, as a function of $r$ in the shower plane,
  after correction of the overall normalization.}
\label{ShapeIndependence}
\end{figure} 

\subsection{Number density of muons at the ground} 
\label{s:number_density}
 
The number of muons per unit area, i.e.\ the muon density $\rho_\mu$, is a two
dimensional function of position coordinates $(x,y)$ relative to the shower
axis position $(x_\text{c},x_\text{c})$, that is
$\vec{r}=(x-x_\text{c},y-y_\text{c})$, projected onto the shower
plane. Descriptions of $\rho_\mu(\vec{r})$ at the ground level have been
obtained and understood with the aid of comprehensive simulations of extensive
air showers~\cite{model,billoir}.

As muons traverse the atmosphere they lose energy by ionization and hard muon
interactions, namely bremsstrahlung, pair production and nuclear interactions
via photo-nuclear processes. Below the critical energy for the muons, which is
of order 500\,GeV, ionization losses dominate.  Decay probability in flight
becomes important by effectively removing muons below an energy threshold
which depends on both the muon energy and the distance traveled~\cite{LCB}. As
the zenith angle increases from $60^\circ$ to $90^\circ$, the distance
traveled to the ground level increases from ${\sim}10$\,km to values over an
order of magnitude larger, and the average energy of the muons at the ground
level rises accordingly. The average energy of muons at the ground level,
produced by a primary hadron at zenith angle $70^\circ$ ($80^\circ$), is about
25 (60)\,GeV.

\begin{figure}[tbp]
\centering
\includegraphics[height=0.43\textwidth]{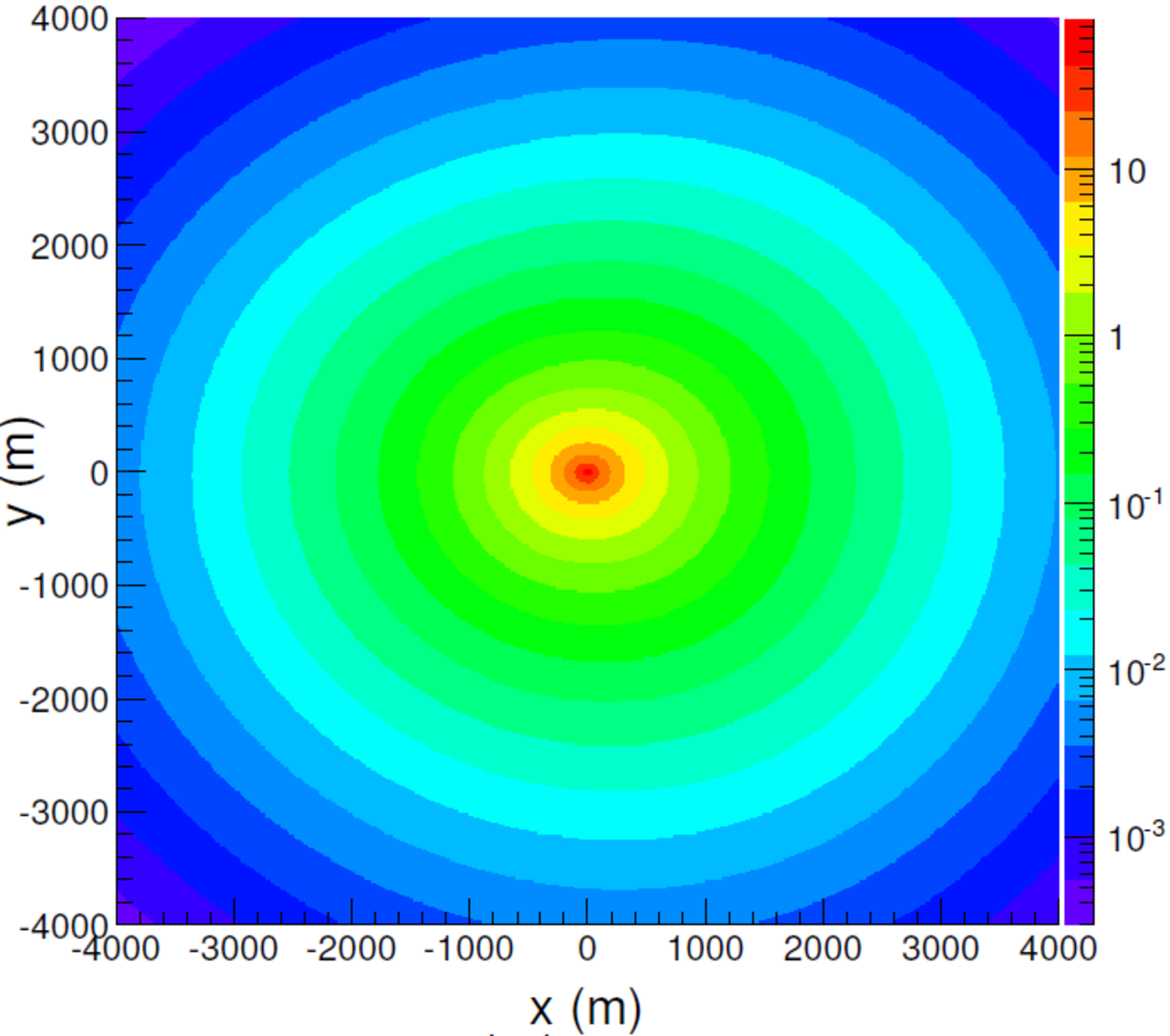}
\hfill
\includegraphics[height=0.415\textwidth]{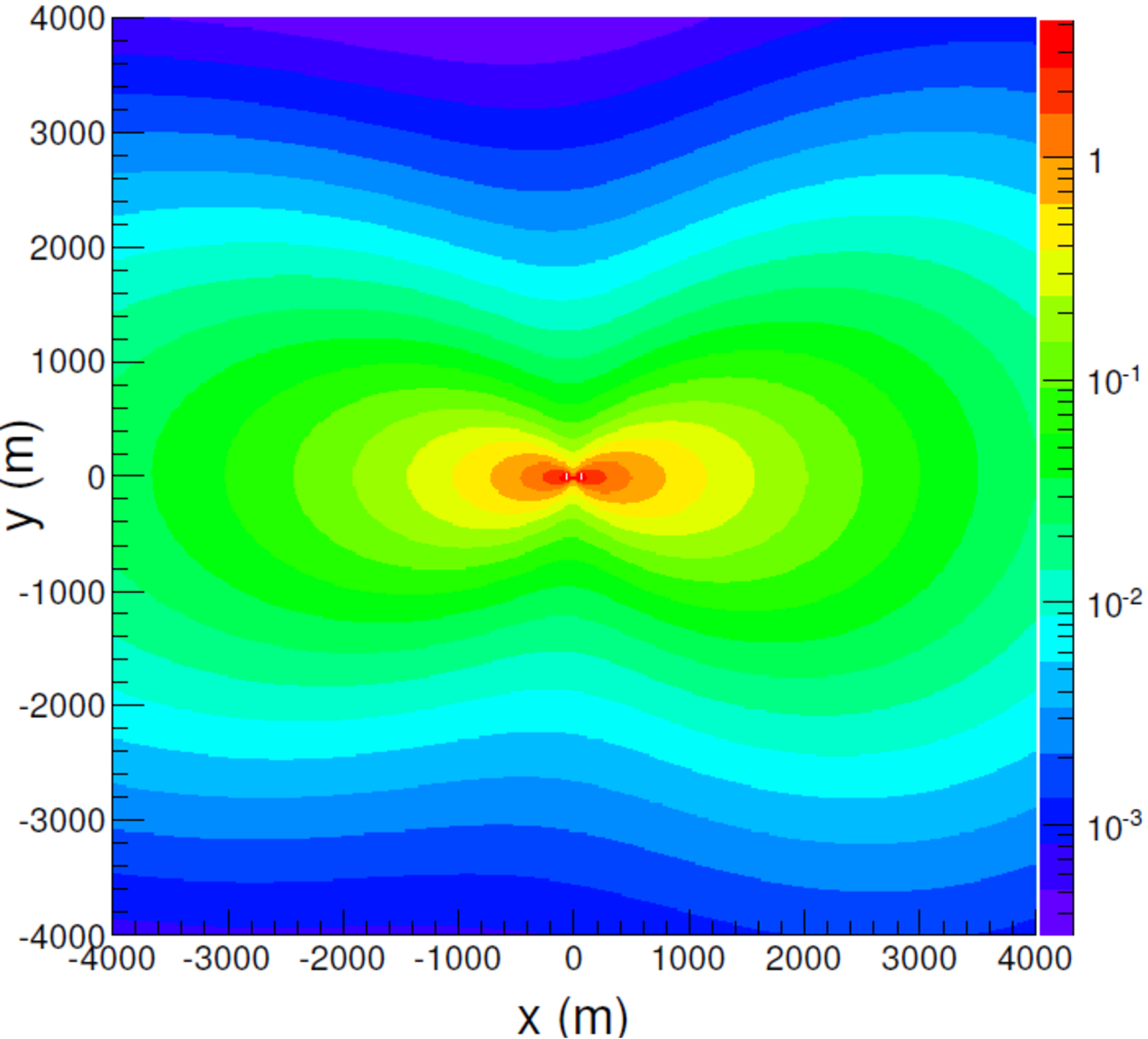}
\caption{Contour plot of the muon density in the shower plane for $E=10$\,EeV
  proton showers with zenith angles $70^\circ$ (left) and $84^\circ$ (right)
  and azimuth angle $0^\circ$, as obtained from simulations based on
  \textsc{QGSJetII-03}. The $y$-axis is oriented in the direction of the $\vec
  B$ field projected onto the shower plane.}
\label{ContourPlot}
\end{figure}

In the Earth's magnetic field $\vec{B}$, the muons are also deflected by the
Lorentz force which laterally separates the positively and negatively charged
muons~\cite{Hillas}. Typical asymmetries are illustrated in
Figs.~\ref{ShapeIndependence} and \ref{ContourPlot}. The separation proceeds
in the direction perpendicular to the plane defined by the shower axis and the
magnetic field.  The magnitude of the separation depends mainly on the
component of $\vec{B}$ perpendicular to the shower direction, the muon energy
and the distance traveled~\cite{model}. The resulting signal patterns at the
ground thus have a quite strong dependency on the arrival directions. As the
zenith angle changes, the large variations in the distance traveled by the
muons are to a large extent responsible for changes in the patterns at the
ground level. As the azimuthal direction of the shower changes, smaller
differences in the patterns are also observed, due to the varying angle
between the typical muon velocity and the $\vec{B}$ field.  Two different
approaches have been used to obtain these distributions. One is based on a
transformation of cylindrically symmetric patterns, exploiting the
anti-correlation between muon energy and angle to the shower
axis~\cite{model}. The other relies on continuous parameterizations in zenith
angle and position in the shower plane that are fitted to results obtained
from simulations~\cite{billoir}.  Both approaches have been shown to reproduce
the average profile of a given set of simulated showers with an accuracy
better than 5\%.

When the arrival direction and the nature of the primary particle are fixed,
the muon number density has been shown to scale nearly linearly with shower
energy ($\rho_\mu \propto E^\alpha$ with $\alpha$ typically being in the range
$[0.90$--$0.95]$)~\cite{model,billoir}. There are some differences between the
distributions depending on the assumed nature of the primary particle, its
energy and the hadronic interaction model used in the simulations. It has also
been shown that these differences are manifested primarily by an overall
normalization of the muon densities, and the shapes of these functions are
approximately the same for a given arrival direction~\cite{Universality},
i.e., weakly dependent on both shower energy and composition.  Both
characteristics are illustrated in figure~\ref{ShapeIndependence}.

The universal shape of the muon distribution and the scaling between muon
number density and shower energy provide the basis of the fitting
procedure. The reconstruction of the shower size is based on the fit of
measured signals to the expected muon patterns. Details are fully described
below.
 
According to the scaling property mentioned above, the expected muon number
density at the ground can be written as:
\begin{equation}
\rho_{\mu}(\vec{r}) = N_{19}\;\rho_{\mu,19}(\vec{r};\theta,\phi).
\label{MDensity}
\end{equation}
Here $N_{19}$ is a measure of the shower size and is the relative
normalization of a particular event with respect to a reference muon
distribution, $\rho_{\mu,19}(\vec{r};\theta,\phi)$, conventionally chosen to
be the average muon density for primary protons with $E=10^{19}$\,eV obtained
with a chosen shower model, \textsc{QGSJetII-03}~\cite{QGSJet2} reference in
our case. The dependence of these functions on the zenith and azimuth angles
$(\theta,\phi)$ is indicated explicitly.

Two sets of muon distributions were generated for comparison purposes, one
following~\cite{model} using the \textsc{Aires}~\cite{AIRES} package for
shower simulations with the \textsc{QGSJet01}~\cite{QGSJet1} model for
hadronic interactions, and the other following~\cite{billoir} with
\textsc{Corsika}~\cite{Corsika} and \textsc{QGSJetII-03}. Examples of such
distributions are shown in figure~\ref{ContourPlot} for different zenith
angles.

The actual value obtained for $N_{19}$ depends upon the particular choice of
composition and hadronic model made for the reference distribution.  The
proton showers obtained with the \textsc{QGSJetII-03} model was chosen as a
reference for the data analysis presented in this work. For instance, the muon
densities for proton showers derived using different high-energy hadronic
interaction models scale with the approximate factors relative to the
reference distribution given in table~\ref{ModelFactors}. A primary
composition different from protons would also enhance the shower size scaling
with the muon content~\cite{NuclearPatterns}. In the extreme case of iron, the
corresponding approximate factors are also given in
table~\ref{ModelFactors}. These factors are indicative of the large expected
uncertainties associated with the hadronic models and the unknown
composition. Nevertheless, these uncertainties do not have a large impact on
the measurement of the energy spectrum where the energy scale had been
inferred from a sub-sample of events measured simultaneously with the FD and
SD (see section~\ref{s:energy_calibration}). This analysis mimics the
procedure to provide an absolute energy calibration~\cite{SpectrumPRL} used
for the reconstruction of events with zenith angle less than $60^\circ$. Most
of the uncertainties associated with the unknown primary composition and
hadronic model, as well as many uncertainties associated with the
reconstruction, are absorbed in this robust and reliable calibration
procedure.

\begin{table}[tbp]
\centering
\begin{tabular}{lll}
\toprule
\emph{Model} & Proton & Iron
\\
\midrule
\textsc{QGSJet01} & 1.10 & 1.46
\\ 
\textsc{QGSJetII-03} & 1.00 & 1.32
\\
\textsc{QGSJetII-04} & 1.19 & 1.58
\\
\textsc{Epos\,1.99} & 1.25 & 1.65
\\
\textsc{Epos\,LHC} & 1.22 & 1.61
\\
\textsc{Sibyll\,2.1} & 0.90 & 1.20
\\
\bottomrule
\end{tabular}
\caption{Scale factors of the muon number densities derived for proton
  and iron showers simulated using different hadronic
  models~\cite{QGSJet1,QGSJet2,QGSJet2-04,EPOS,EPOS-LHC,SIBYLL}, relative to
  the reference muon distribution based on protons using \textsc{QGSJetII-03}.}
\label{ModelFactors}
\end{table}

\subsection{Electromagnetic component}
\label{sec_emcomponent}

As inclined showers develop in the atmosphere, the EM cascade produced by the
neutral pions in the hadronic collisions is attenuated, leaving a front
dominated by muons. The remaining EM component (electrons, positrons and
photons) originates from two different mechanisms. First is due to the tail of
the hadronic cascade and decreases very rapidly as the zenith angle
increases. Its magnitude depends on composition and interaction models. This
component increases with the primary energy and varies according to the
fluctuations in shower maximum. This dependence has to be considered as a
source of systematic uncertainty in the electromagnetic correction, as will be
discussed in section~\ref{sec_sys}. The second component is produced by the
muons themselves and closely follows the muon density patterns. It is mainly
due to the showering of the electrons from muon decay in flight, although
high-energy muons very close to the core also contribute through pair
production and bremsstrahlung. More details can be found in~\cite{Ines}.

In the reconstruction procedure (see section 4) the signals measured with the
SD are compared to the reference muon distribution, and for this purpose it is
necessary to extract the signal induced by muons from the total signal at each
detector. This is achieved by subtracting the EM component from the detector
signal using the average ratio, $R_{\text{EM}/\mu}(r,\theta)$, of the
electromagnetic to the muonic contributions:
\begin{equation}
R_{\text{EM}/\mu}(r,\theta) = S_\text{EM}(r,\theta)/S_\mu(r,\theta).
\label{EMratio}
\end{equation}
This ratio was studied using Monte-Carlo simulations. Parameterizations of the
average electromagnetic ($S_\text{EM}$) and muonic ($S_\mu$) contributions to
the signal were obtained separately in terms of distance to the shower axis
$r$, and of zenith angle $\theta$.  Proton simulations at $10^{19}$\,eV with
\textsc{QGSJet01} was chosen as a reference. The accuracy of this
parameterization is better than 5\%.

\begin{figure}[tbp]
\centering
\includegraphics[width=0.48\textwidth]{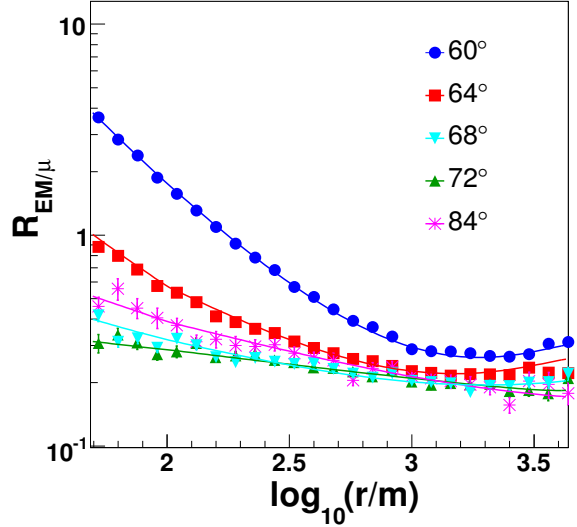}
\caption{Ratio of the electromagnetic to muonic contributions to the detector
  signal averaged over the polar angle as a function of $r$, at different
  zenith angles.}
\label{EMCorrection}
\end{figure}

In figure~\ref{EMCorrection} the ratio $R_{\text{EM}/\mu}$, averaged over the
polar angle (with respect to the shower axis projected onto the shower plane),
is shown as a function of $r$ for different $\theta$. The ratio can be seen to
decrease as $\theta$ increases from ${\sim}60^\circ$ to $68^\circ$. At larger
$\theta$ and near the shower axis, $R_{\text{EM}/\mu}$ can be seen to increase
slightly due to hard muon interaction processes. At distances from the shower
axis exceeding 1\,km, $R_{\text{EM}/\mu}$ changes only weakly with distance,
typically lying between 15\% and 30\%.

\begin{figure}[tbp]
\centering
\includegraphics[width=0.42\textwidth]{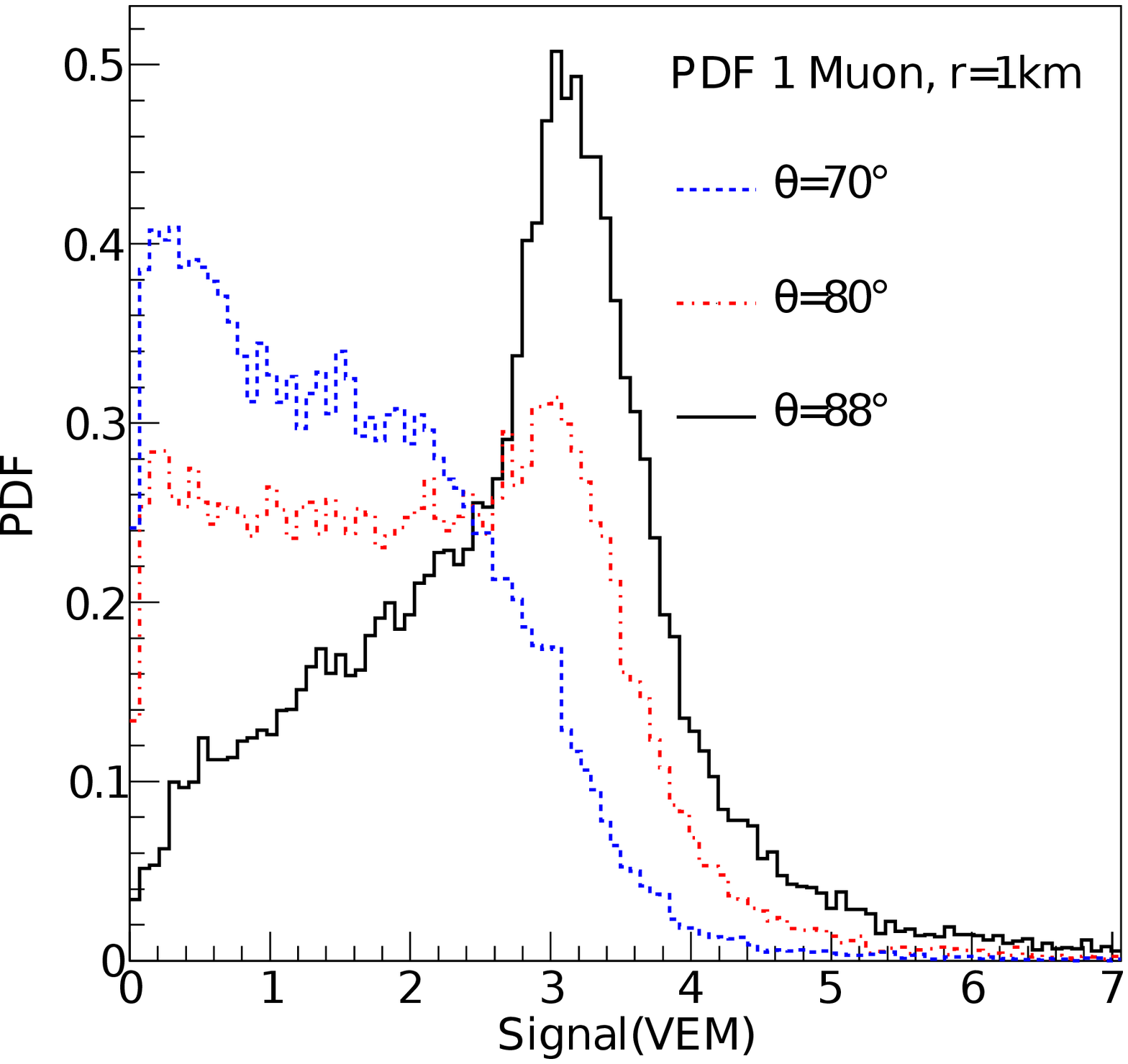}
\hfill
\includegraphics[width=0.42\textwidth]{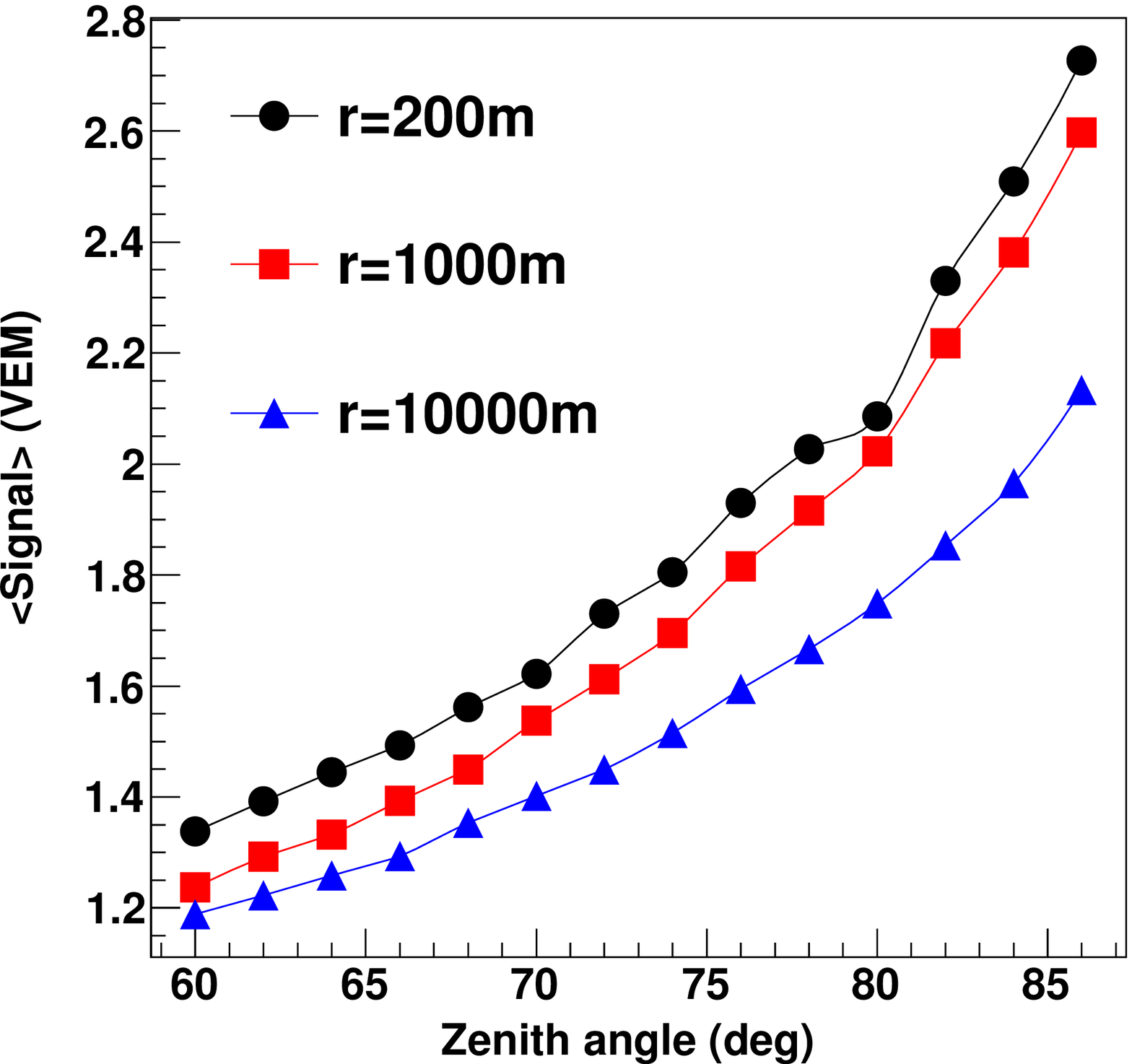}
\caption{\emph{Left:} probability distribution of the signal produced by
  single muons at a fixed distance of 1000\,m from the shower axis, and for
  three different zenith angles. \emph{Right:} average values of the signal
  produced by single muons as a function of zenith angle, for three different
  distances, as labeled.}
\label{SignalvsTheta}
\end{figure} 

\subsection{Signal response of the surface detector stations}

To reconstruct the position of the shower core and shower size $N_{19}$, the
measured signals are fitted to the model predictions. The fit requires the
evaluation of the probability density function (PDF) of the signal response of
each surface detector to the expected number of muons. The PDF distributions
are functions of the signal $S_\mu^\text{meas}$ deposited by muons in the
detector, which is in turn estimated from the measured signal $S^\text{meas}$,
using the average ratio of electromagnetic-to-muonic signals described in the
previous section (eq.~\eqref{EMratio}):
\begin{equation}
S_\mu^\text{meas}=\frac{S^\text{meas}}{1 + R_{\text{EM}/\mu}}.
\label{Smu}
\end{equation}
The PDF for each surface detector is constructed from the expected number of
muons, assuming Poisson statistics and using a simpler PDF corresponding to
the passage of a fixed number of muons.

\begin{figure}[tbp]
\centering
\includegraphics[width=0.41\textwidth]{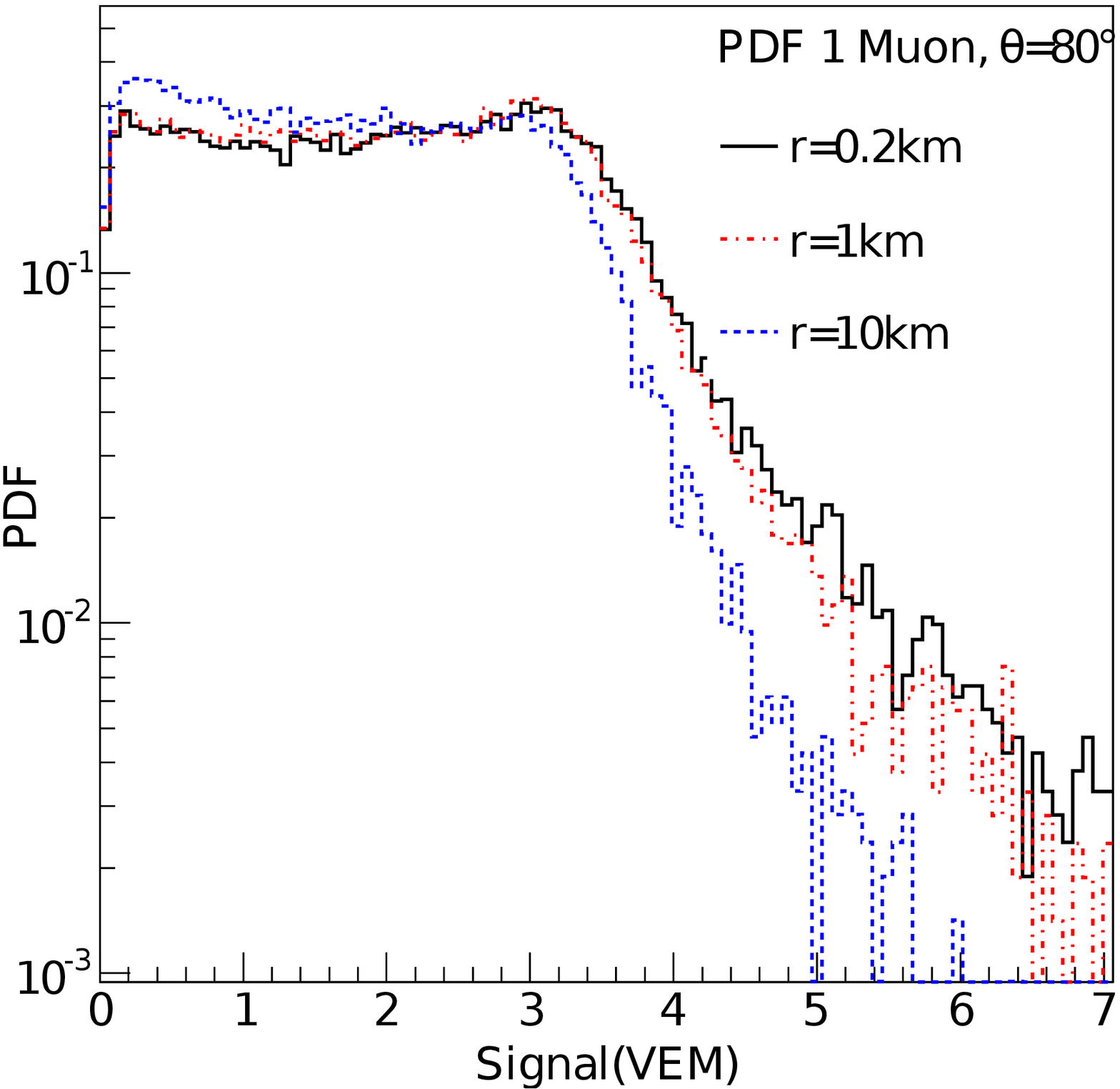}
\hfill
\includegraphics[width=0.42\textwidth,clip=]{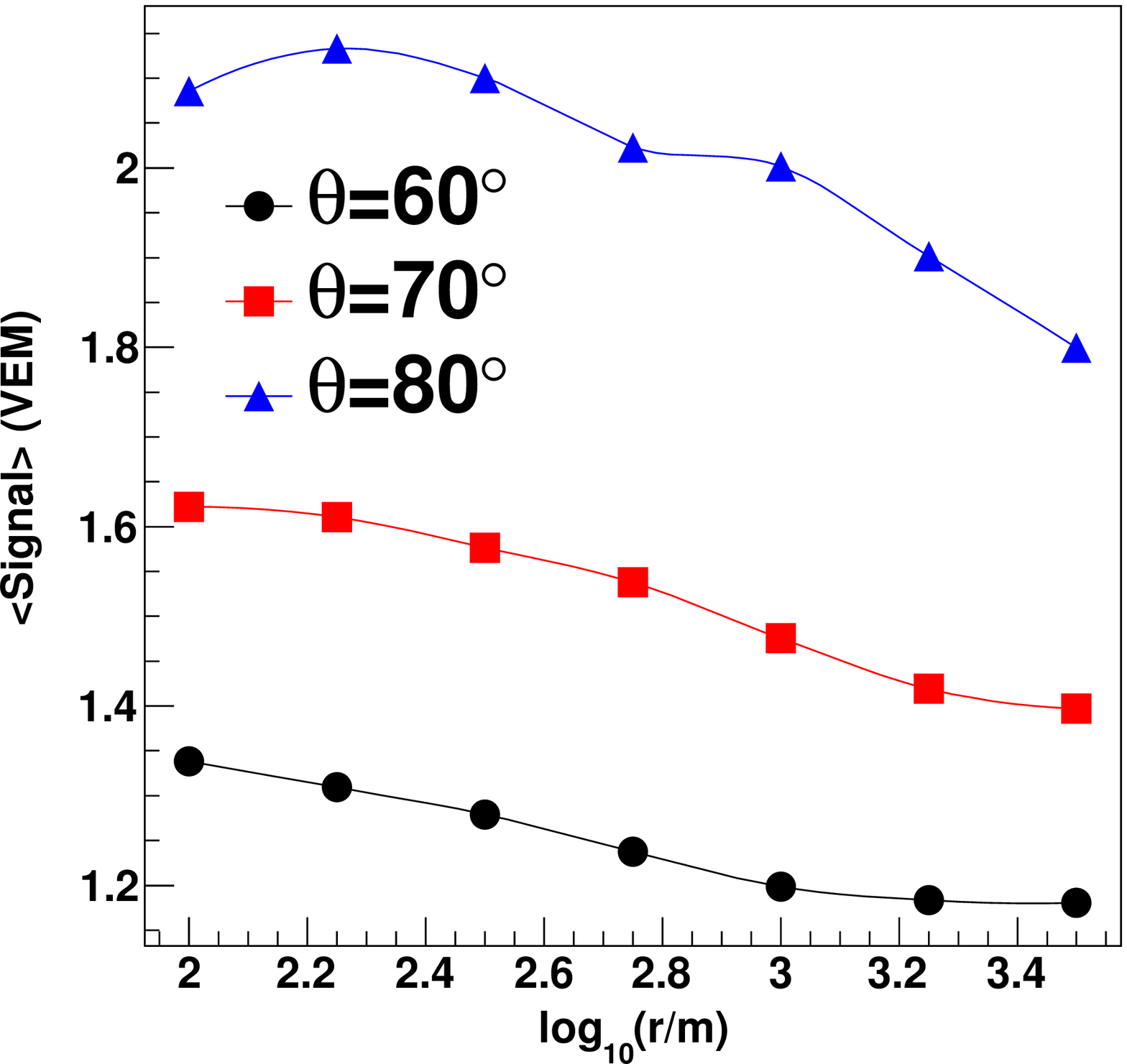}
\caption{\emph{Left:} probability distribution of the signal produced by
  single muons of $80^\circ$ and three different distances $r$ to shower
  core. \emph{Right:} average signal produced by single muons as a function of
  $r$, for three zenith angles, as labeled.}
\label{SignalvsEnergy}
\end{figure}

The basic prerequisite for these probability densities is the signal
distribution for a detector hit by a single muon. These were obtained with
high statistics using a module in the official software framework
\mbox{$\overline{\textrm{Off}}$\hspace{.05em}\protect\raisebox{.4ex}{$\protect\underline{\textrm{line}}$}}~\cite{Offline}
of the Pierre Auger Observatory, which simulates the detector response and
interfaces it with the \textsc{Geant4} package~\cite{GEANT}. Multiple
histograms for the signal response were generated for discrete values of
$\theta$, and different relative positions of the station with respect to the
shower core~\cite{GRFThesis}. They are illustrated in
Figs.~\ref{SignalvsTheta} and \ref{SignalvsEnergy}. The signal is mainly due
to the Cherenkov light emission by the muon tracks, and the PDF is closely
related to the track-length distributions of the muons inside the detectors,
which have a strong dependence on the zenith angle, as explicitly shown in
figure~\ref{SignalvsTheta}.

The simulations include also contributions that do not scale with the muon
track length and are dependent on the relative position of the stations. Delta
rays (i.e., scattered electrons within the detector) account for an
enhancement of the signal of order 20\%, which increases with
energy~\cite{Ave}. For high-energy muons bremsstrahlung, pair production and
nuclear interactions inside the detector~\cite{Ave} give contributions which
appear as harder tails of the response functions for stations close to the
shower core. Low-energy muons have a reduced Cherenkov efficiency (enhanced by
energy loss), becoming zero at the Cherenkov threshold. The
average signal increases as the distance to the shower axis is reduced (as
illustrated in figure~\ref{SignalvsEnergy}) due to the increase in muon energy.
The contribution, to the average signal, of the direct Cherenkov light that
hits the photomultipliers without any reflection from the walls of the
detector, ranges from 3\% at $\theta=60^\circ$ to 10\% at $80^\circ$.

\begin{figure}[tbp]
\centering
\includegraphics[width=0.48\textwidth]{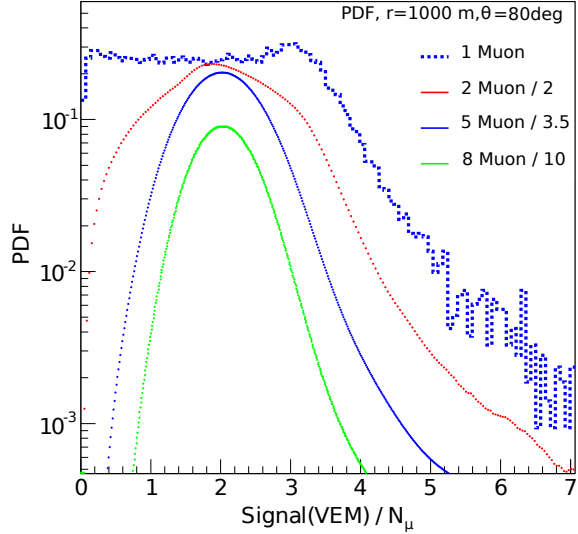}
\caption{Probability distribution of the signal for a single muon with a
  zenith angle of $80^\circ$, falling 1000\,m from the shower core, compared
  to the convolution for 2, 5 and 8 such muons. The distributions have been
  rescaled by a convenience factor given in labels.}
\label{SeveralMuons}
\end{figure}

Response histograms for single muons are used to obtain the response of the
detector to the passage of multiple ($k$) muons by convolution in an iterative
fashion, when $k$ is between 1 and 8. When $k>8$ a Gaussian approximation is
used with an average $\xi_k=k\,\xi_1$ and a standard deviation of
$\sigma_k=\sqrt{k}\,\sigma_1$, in terms of the averages ($\xi_1$) and standard
deviations ($\sigma_1$) of the corresponding single muon histograms.  As the
number of muons grows, histograms rapidly become Gaussian-like, as anticipated
from the Central Limit Theorem, see Fig~\ref{SeveralMuons}. We label these PDF
distributions as:
\begin{equation} 
p_\text{st}(S_\mu^\text{meas}; k,\vec{r},\theta),
\label{pdf}
\end{equation}
explicitly indicating the dependence on $k$, $\theta$ and $\vec r$. The
statistical uncertainty of these histograms is below $5\%$. A total of 12\,960
histograms is needed to describe the signal distributions for the passage of
$k$ muons ($k\,\le\,8$) in 15 bins of $\theta$, 12 bins of polar angle and 9
bins of distance $r$.

\section{Reconstruction}
\label{s:reconstruction}

The reconstruction of inclined showers is performed first by fitting the
arrival directions of the events using the start times of the signals, and
then fitting the muon density patterns at the ground level to the signals
measured by the surface detectors.

The reconstruction was also validated using large samples of simulated
events. Extensive libraries of isotropically arriving proton and iron showers
were made with a thinning level~\cite{thinning} of $10^{-6}$.  A set of events
simulated with \textsc{Aires} and \textsc{QGSJet01} and a spectral index of
$\gamma=2.6$ in the energy range $\log_{10}(E/\text{eV})=[18.5,20]$, and a
zenith angle between $50^\circ$ and $89^\circ$ (100\,000 events each) was
used. A second and third set were simulated with \textsc{Corsika} using
\textsc{QGSJet-04} or \textsc{Epos\,LHC}, with a spectral index of $\gamma=1$
between $60^\circ$ and $89^\circ$.  Each of these sets covers proton and iron
in the same energy interval and is further split into three equal subintervals
in $\log_{10}E$ (each energy subinterval has 30\,000 events).  Further
libraries were also generated with \textsc{QGSJetII-03} and
\textsc{Epos\,1.99}. All of these showers have subsequently undergone a full
simulation of the detector with random impact points in the SD array, within
the
\mbox{$\overline{\textrm{Off}}$\hspace{.05em}\protect\raisebox{.4ex}{$\protect\underline{\textrm{line}}$}}
framework, to generate databases of simulated events.

Unless otherwise indicated, the criteria used to select the simulated events
for comparison match the selection that will be used for the measurement of
the spectrum with inclined data. Events with $60^\circ<\theta<80^\circ$ are
required to pass the inclined T4 and T5 conditions, with also $N_{19}>0.7$ to
ensure the SD array is fully efficient.

\subsection{Angular reconstruction}
\label{ss:angularreconstruction}

The directions of the incident cosmic rays are determined from the relative
arrival times of the shower front in the triggered stations. The start times
of the signals are fitted to those expected from a shower front with
curvature, using a standard $\chi^2$ minimization procedure.

The angular accuracy depends on the number of triggered stations, on sizes of
their signals, and also on the zenith angle of the shower itself.  Typically,
as the shower becomes more inclined, the arrival direction is reconstructed
better, although this trend is inverted for events above $80^\circ$. As the
shower becomes dominated by the muons, its front develops a smaller time
spread than for vertical showers, and the start time can be established more
precisely. In addition, as the zenith angle increases, the curvature of the
shower front is reduced since the muons are produced further away from the
ground and have higher energies. Finally, the number of triggered stations in
a shower tends to increase as the zenith angle rises, due simply to the
projection of the station positions into the shower plane.

Equivalent precision on angular reconstruction has been obtained with the
different models used to describe the shower front and the variance reported
in~\cite{LCB,CBB}.

\begin{figure}[tbp]
\centering
\includegraphics[height=0.36\textwidth]{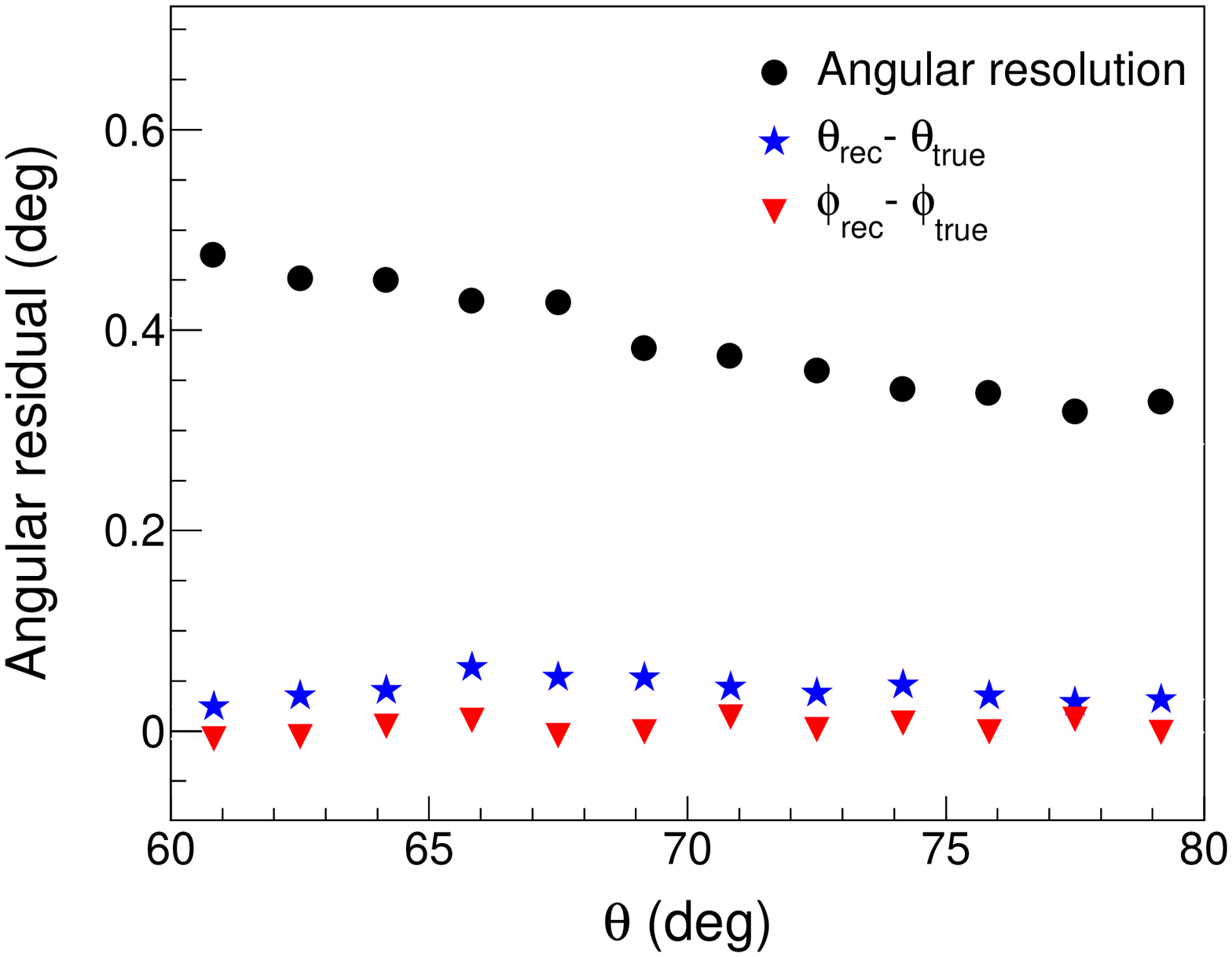}
\hfill
\includegraphics[height=0.36\textwidth]{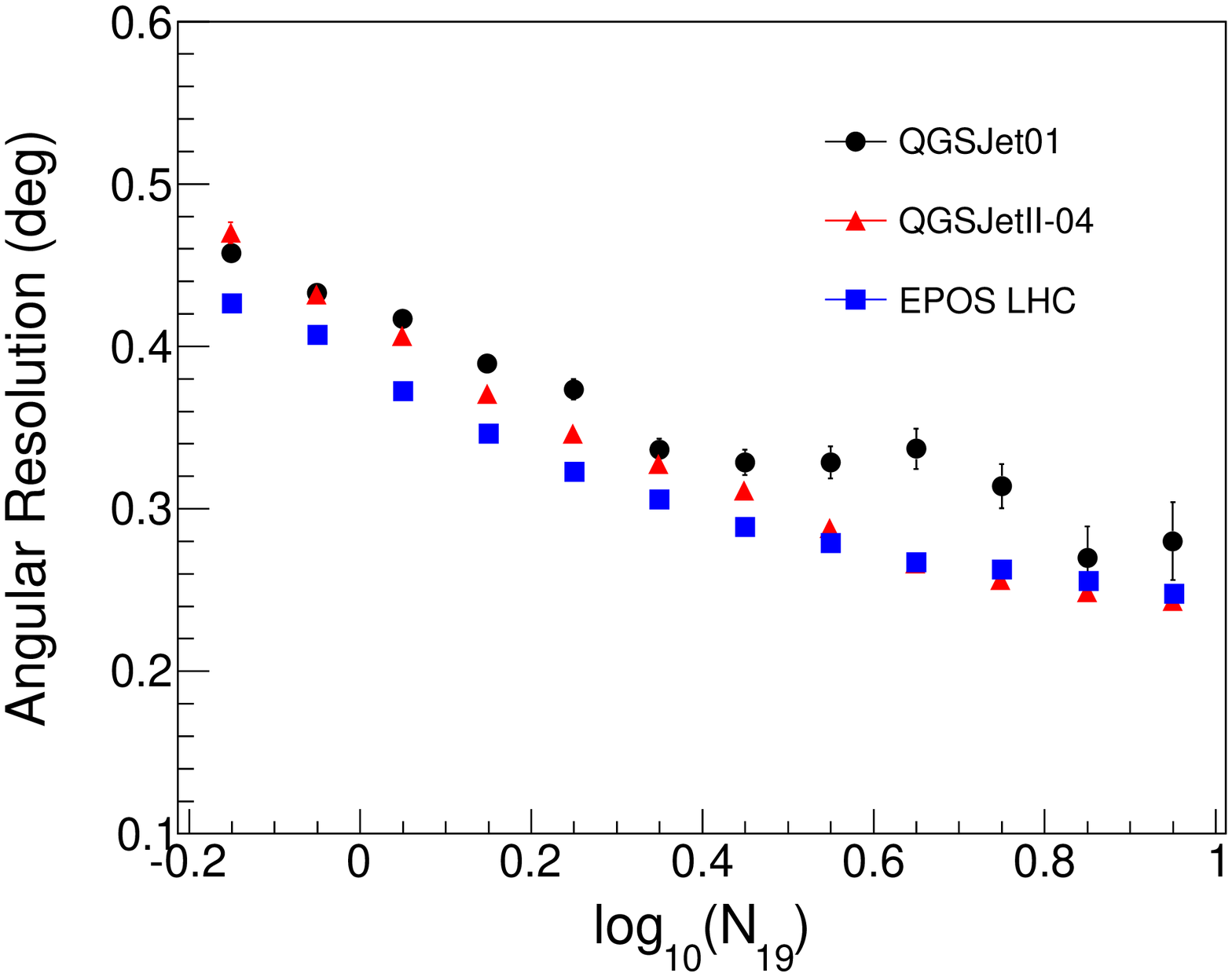}
\caption{Comparison of reconstructed and true arrival directions with
  simulated events. \emph{Left:} difference between true and reconstructed
  zenith (blue stars) and azimuth (red triangles) angles, as a function of the
  true zenith angle, for proton SD events simulated with
  \textsc{QGSJet01}. The angular resolution (black circles) is also
  shown. \emph{Right:} angular resolution as a function of shower size for
  different hadronic models, \textsc{QGSJet01} (black circles),
  \textsc{QGSJetII-04} (red triangles) and \textsc{Epos\,LHC} (blue squares).}
\label{DeltaAngle}
\end{figure} 

The angular reconstruction was tested with simulations. The left panel of
figure~\ref{DeltaAngle} displays the difference between true and reconstructed
angles, as well as the angular resolution as a function of zenith angle for
showers simulated with \textsc{QGSJet01} and for model~\cite{LCB}. The angular
resolution, defined as the angle at 68\% of the cumulative distribution
function of the space angle (angular separation between the true and
reconstructed directions), is obtained by fitting this distribution with a
Gaussian resolution function, $\mathrm{d}p\propto
\exp(-\alpha^2/2\sigma^2)\,\mathrm{d}\cos\alpha\,\mathrm{d}\phi$, and
corresponds to $1.5\times\sigma$~\cite{AngularResolution}.  The right panel of
figure~\ref{DeltaAngle} displays the angular resolution as a function of
shower size, comparing the performance using different shower simulations and
hadronic models.  The reconstructed zenith and azimuth angles have a bias less
than $0.08^\circ$ and $0.02^\circ$, respectively, and the angular resolution
better than $0.5^\circ$, improving to better than $0.35^\circ$ for the highest
energies.

The angular reconstruction of the SD was studied using hybrid data, data from
a region of the array with stations on a 750\,m grid, events with stations
that are duplicated at given positions, and different samples of simulated SD
events. These approaches have yielded compatible results.

\subsection{Shower size and core position reconstruction}

\subsubsection{Procedure}
\label{ss:recprocedure}

Once the arrival direction is established, the expected number of muons
$n_\mu$ at each station can be obtained multiplying the corresponding muon
number density (eq.~\eqref{MDensity}) by the detector area $A_\perp$ projected
onto the shower plane:
\begin{equation}
n_\mu =
  \rho_{\mu}(\vec{r})\;
  A_\perp(\theta) =
    N_{19}\;
    \rho_{\mu,19}(\vec{r};\theta,\phi)\;
    A_\perp(\theta).
\label{eq::meanmu}
\end{equation}

Estimates of the position of the shower core $(x_\text{c},y_\text{c})$ and the
shower size $N_{19}$ are obtained by fitting the expected number of muons
$n_\mu$ to the muonic part of the measured signal $S_\mu^\text{meas}$
(eq.~\eqref{Smu}).  The fitting is performed using a maximum-likelihood method
including the information from non-triggered and saturated stations. This
constrains the shower size and reduces any selection bias due to the threshold
trigger. The log-likelihood function is the logarithm of the combined
probability $p$ to obtain the measured signals in all detectors.  For each
triggered and non-triggered participating station, this is taken as the
product of the probability densities of getting the muonic signal
$S_\mu^\text{meas}$ when $n_\mu$ muons are expected. Therefore, the
log-likelihood function is given by:
\begin{equation}
\log\mathcal{L} = \sum_{i=1}^N\log p_i(S_\mu^\text{meas};n_\mu,\vec{r},\theta) 
\label{eq::likelihood}
\end{equation}
and depends on the three free parameters which enter the calculation of
$n_\mu$ (eq.~\eqref{eq::meanmu}), two for the shower core position
$(x_\text{c},y_\text{c})$ and one for the shower size $N_{19}$.
 
\begin{figure}[tbp]
\centering 
\includegraphics[width=0.5\textwidth]{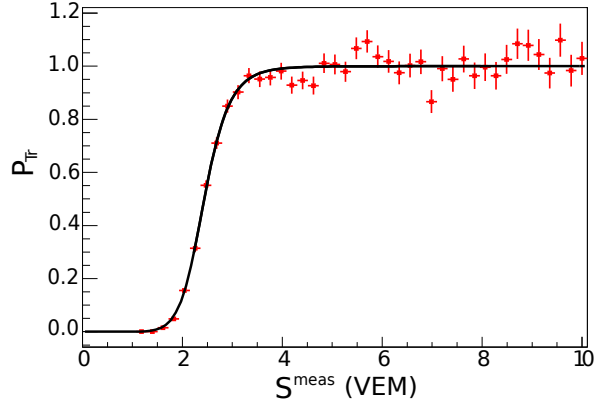}
\caption{Estimate of the average probability that a detector triggers in terms
  of the measured signal.}
\label{ThresholdProbability}
\end{figure}

The measured signal $S_\mu^\text{meas}$ when $n_\mu$ muons are expected, can
be produced by different number of muons $k$, each with different probability
density function $p_\text{st}$ (see eq.~\eqref{pdf} and
figure~\ref{SeveralMuons}). To compute each local probability $p$, the sum
over all the probabilities $p_\text{st}$ for all possible numbers of muons $k$
to produce the measured signal has to be computed.  Each probability must be
weighted by the Poisson probability $\mathcal{P}_\text{oisson}(k;n_\mu)$ of
$k$ muons entering the station when $n_\mu$ are expected. Then, the overall
probability of obtaining $S_\mu^\text{meas}$ in a given detector becomes:
\begin{equation}
p(S_\mu^\text{meas};n_\mu,\vec{r},\theta) =
  P_\text{tr}(S^\text{meas})\,
  \sum_{k=1}^\infty\mathcal{P}_\text{oisson}(k;n_\mu)\;
                       p_\text{st}(S_\mu^\text{meas};k,\vec{r},\theta),
\label{Likelihood}
\end{equation}
where the average trigger probability $P_\text{tr}(S^\text{meas})$ in terms of
the signal at each station is also included. $P_\text{tr}(S^\text{meas})$,
estimated using the signal distributions of the triggered stations in the
data, is shown in figure~\ref{ThresholdProbability}.

The sum naturally accounts for Poisson fluctuations in the number of muons but
additional signal fluctuations enter through $p_\text{st}$. The infinite sum
is truncated for practical reasons.

\begin{figure}[tbp]
\centering
\includegraphics[width=0.43\textwidth]{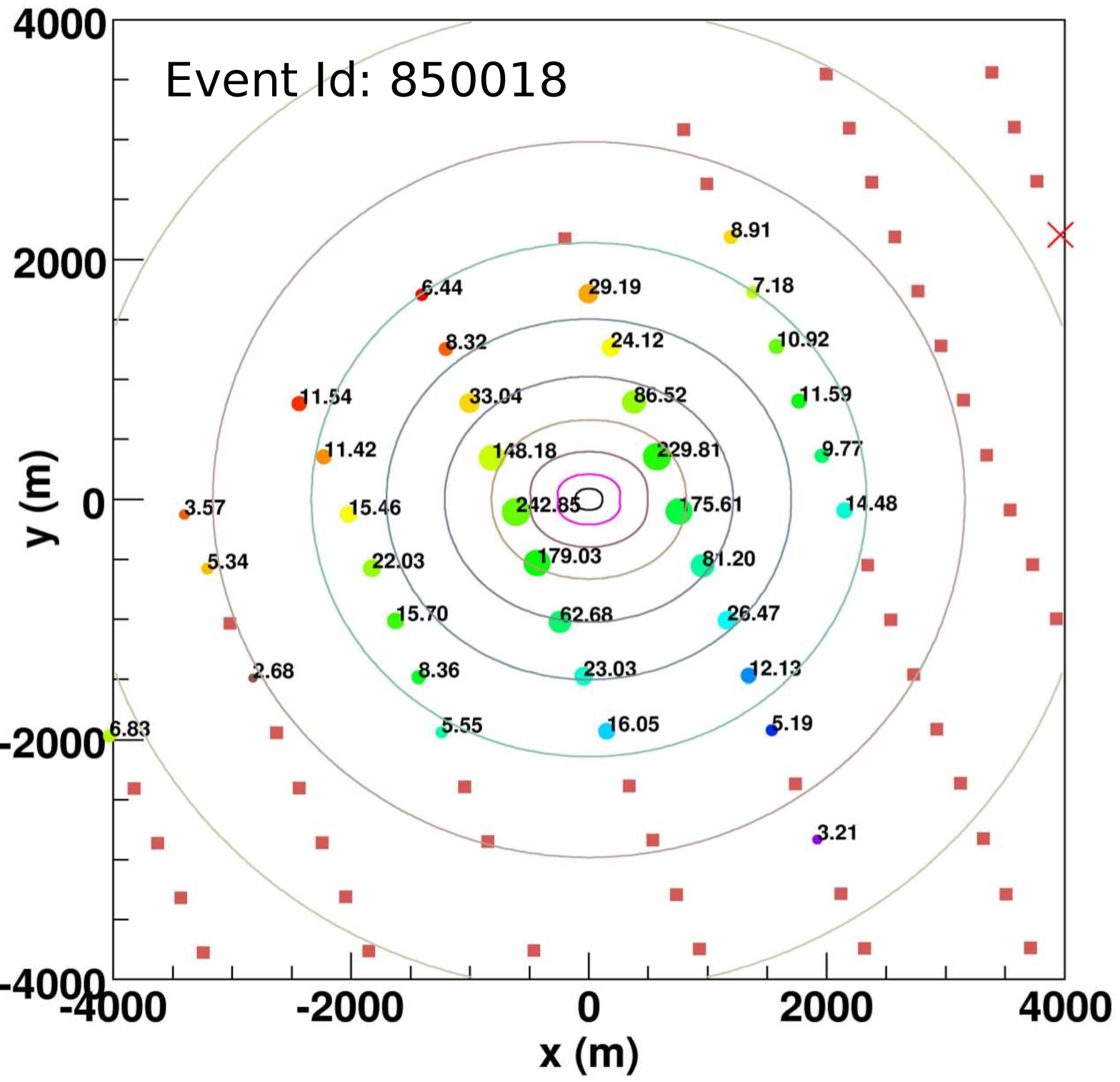}
\hfill
\includegraphics[width=0.51\textwidth]{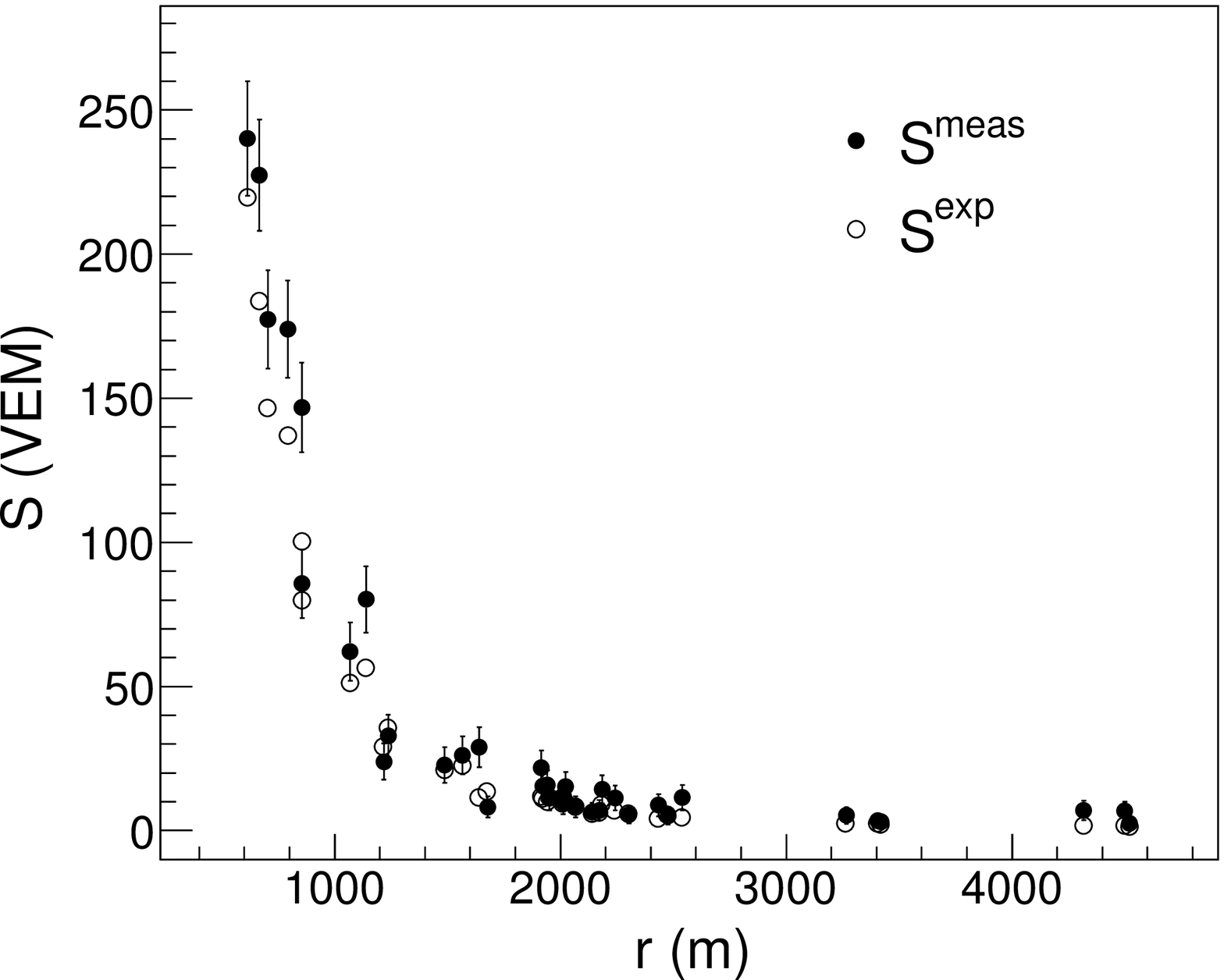}
\caption{\emph{Left:} display of a particular event projected onto the shower
  plane with the contour plot of the fitted distribution superimposed,
  indicating the signal measured in the 37 triggered stations and the position
  of the reconstructed core. The reconstructed zenith angle is $71^\circ$ and
  the best fit value of $N_{19}$ is 9.2 which corresponds to an energy of
  54.6\,EeV after the calibration procedure. The color code indicates the
  start time from early (blue) to late (red) stations. \emph{Right:} signal
  sizes in the triggered stations as a function of the distance to the shower
  core in the shower plane for the same event. Filled and open symbols indicate measured and
  expected signals, respectively.}
\label{Event}
\end{figure} 

For stations that have no signal the probability distribution has to be
replaced by the probability that the detector does not trigger. The
probability that a station does not trigger can be obtained by integrating the
detector response functions $p_\text{st}$, weighted by the probability
function $1-P_\text{tr}(S^\text{meas})$ for not triggering. This is in turn
obtained by summing the Poisson probabilities of having $k$ muons traversing
the station, weighted by the corresponding probabilities for not triggering:
\begin{equation}
p(S_\mu^\text{meas}=0,n_\mu;\vec{r},\theta) = 
  \sum_{k=0}^\infty\mathcal{P}_\text{oisson}(k;n_\mu) 
  \int_0^\infty(1-P_\text{tr}(S))\,
               p_\text{st}(S;k,\vec{r},\theta)\,
               \mathrm{d}S.
\nonumber
\label{ModifySilentWeight}
\end{equation}

To limit the computing time of the minimization procedure, only detector
stations that have no signal and are within four concentric hexagons around
all the triggered stations are included in the likelihood maximization
procedure. For instance, the number of non-triggering stations entering the
fit is on average ${\sim}8$ times the number of stations with signal. The
number of stations without signal used for the reconstruction has a minor
effect on the best fit value of $N_{19}$, and becomes negligible when
sufficient stations have been considered. The average change in shower sizes
when switching from four to two hexagons are below 5\%.

\begin{figure}[tbp]
\centering
\includegraphics[width=0.48\textwidth]{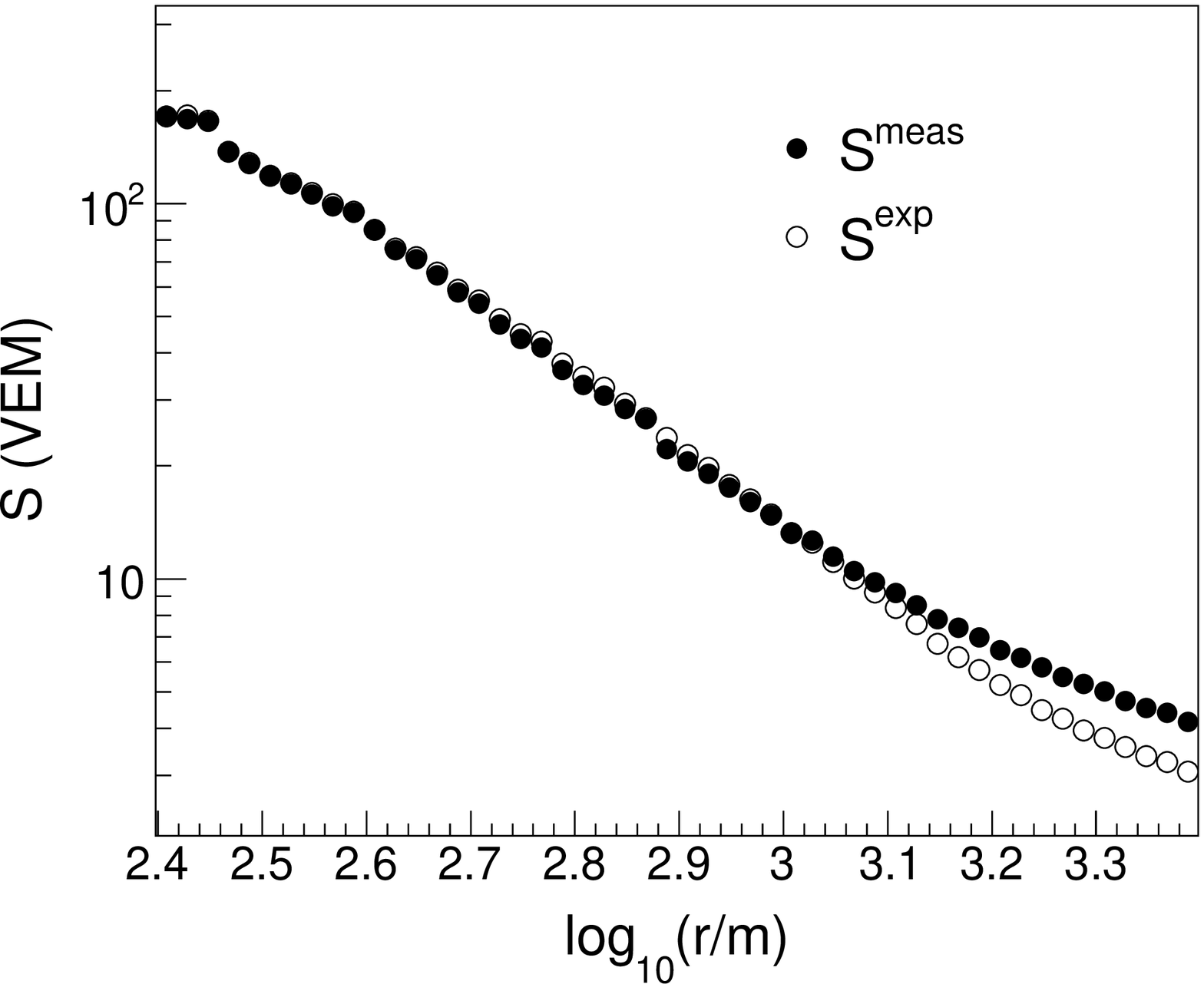}
\hfill
\includegraphics[width=0.48\textwidth]{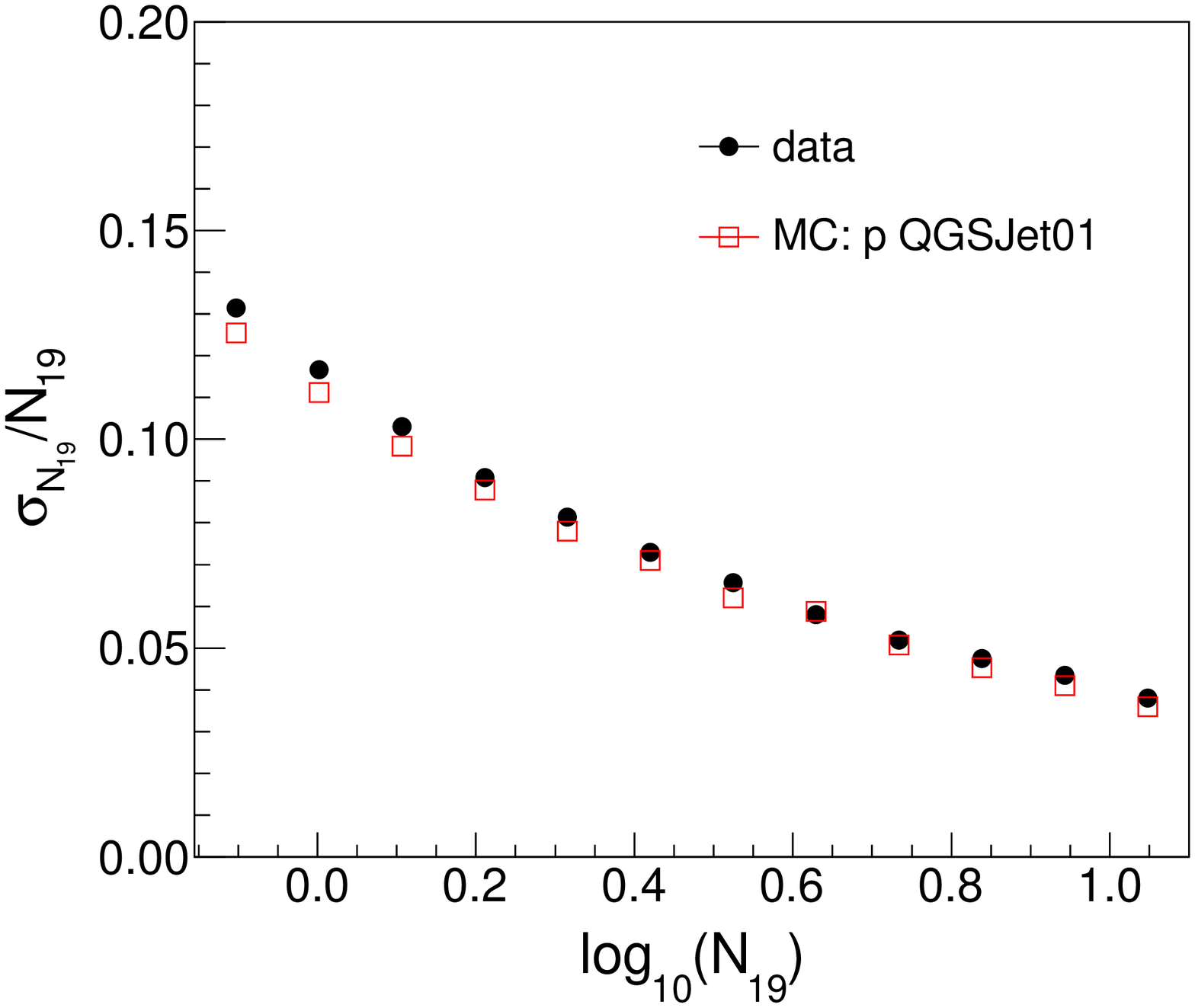}
\caption{\emph{Left:} signal size as a function of the distance to the shower
  core in the shower plane, for proton events simulated with
  \textsc{QGSJet01}. Filled and open symbols indicate measured and expected
  signals, respectively. \emph{Right:} the average statistical relative
  uncertainty in $N_{19}$ as a function of $\log_{10}N_{19}$ for data and
  simulated events.}
\label{Sbias}
\end{figure} 

\subsubsection{Performance}
\label{ss:recperformance}

An example of a reconstructed event from the data is shown in the left panel of
figure~\ref{Event} indicating the position of the shower core and the contour
plots of the average muon number density, in relation to the measured signals
in the stations. The expected muon densities can be obtained by multiplying
the density of the model distribution by the reconstructed value of
$N_{19}$. The values of the expected and measured signals are shown in the
right panel of figure~\ref{Event} for the same event. For each triggered
station, the expected muonic signal $S_\mu^\text{exp}$ is obtained by summing
over all the average signal responses for all possible numbers of muons $k$
with Poisson weights, $\mathcal{P}_\text{oisson}(k;n_\mu)$. Then, the expected
signal $S^\text{exp}$ is estimated from $S_\mu^\text{exp}$ using the average
ratio of electromagnetic and muonic signals: $S^\text{exp}= S_\mu^\text{exp}
(1 + R_{\text{EM}/\mu})$ (analogous to eq.~\eqref{Smu}). 

The performance is evaluated by studying the reconstruction of events obtained
with full simulation. In the left panel of figure~\ref{Sbias} the expected and
measured signals are compared for a sample of simulated events. The
reconstruction procedure achieves an agreement between expected and measured
signals at the 5\% level for expected signals above 10\,VEM. Below this
number, the comparison is not relevant due to the trigger effects and
associated upward fluctuations of the signal.  This is the reason why the
``non-triggered'' stations should be considered in the reconstruction, as
explained in the previous section~\ref{ss:recprocedure}.

The reconstruction of the core position depends on the zenith angle,
nevertheless this dependency is reduced when converted to the shower
plane. The average distance between the reconstructed and true core positions
in this plane is 108\,m, ranging between 80 to 160\,m as the zenith angle
increases from $60^\circ$ to $80^\circ$.

The relative uncertainty in $N_{19}$ as obtained from the fit is shown in the
right panel of figure~\ref{Sbias} for data and simulated events. It decreases
from about 13\% at $N_{19}=0.7$ ($E=4$\,EeV) to about $4\%$ at $N_{19}=10$
($E=60$\,EeV).

\begin{figure}[tbp]
\centering 
\includegraphics[width=0.5\textwidth]{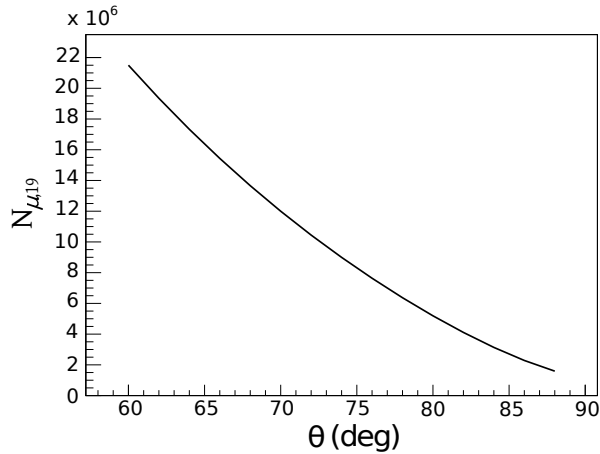}
\caption{Total number of muons in the reference distribution as a function of
  the shower zenith angle. The reference distribution is based on proton
  showers with the energy of $10^{19}$\,eV simulated using the
  \textsc{QGSJetII-03} hadronic interaction model.}
\label{Nmu19}
\end{figure}

An approximate estimate of the true value of $N_{19}$ can be made for each
simulated shower and compared to the reconstructed value. The number of muons
$N_\mu$ that have reached the ground is obtained and normalized to the number
of muons $N_{\mu,19}$ in the reference distribution (i.e.\ the integral of
$\rho_{\mu,19}$ in eq.~\eqref{MDensity} over an area),
\begin{equation}
R_\mu=\frac{N_\mu}{N_{\mu,19}}.
\label{R_mu}
\end{equation}
The total number of muons in the reference distribution is shown in
figure~\ref{Nmu19} where it is clear that the muon number is attenuated with
the zenith angle, whereas, by definition, $N_{19}$ is independent of the
zenith angle. The muon number is essentially independent of azimuth angle. For
a given particle species and arrival direction the ratio, $R_\mu$, scales with
shower energy, on average, and is normalized to $10^{19}$\,eV, so that it can
be compared directly with $N_{19}$.

\begin{figure}[tbp]
\centering
\includegraphics[height=0.34\textwidth]{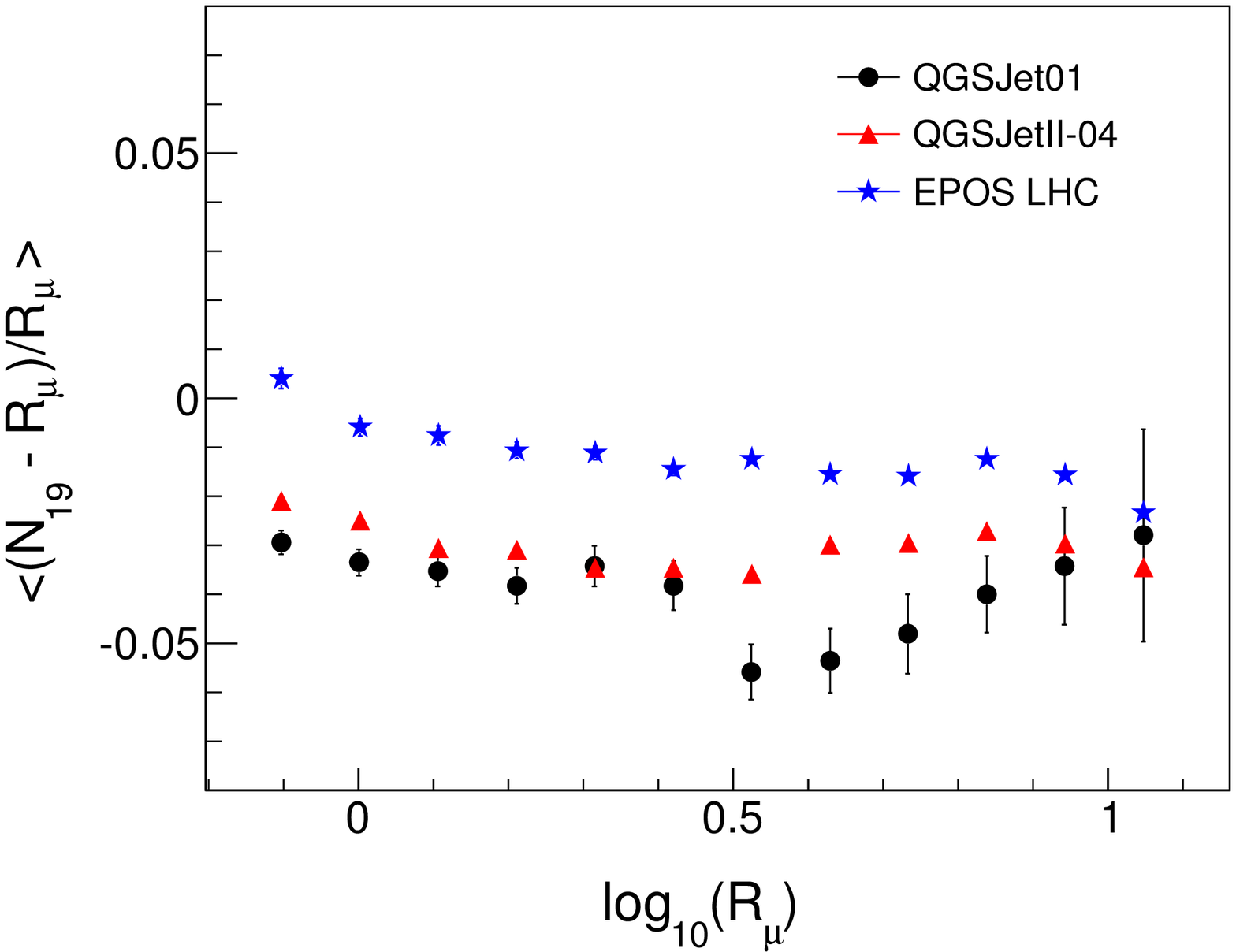}
\hfill
\includegraphics[height=0.34\textwidth]{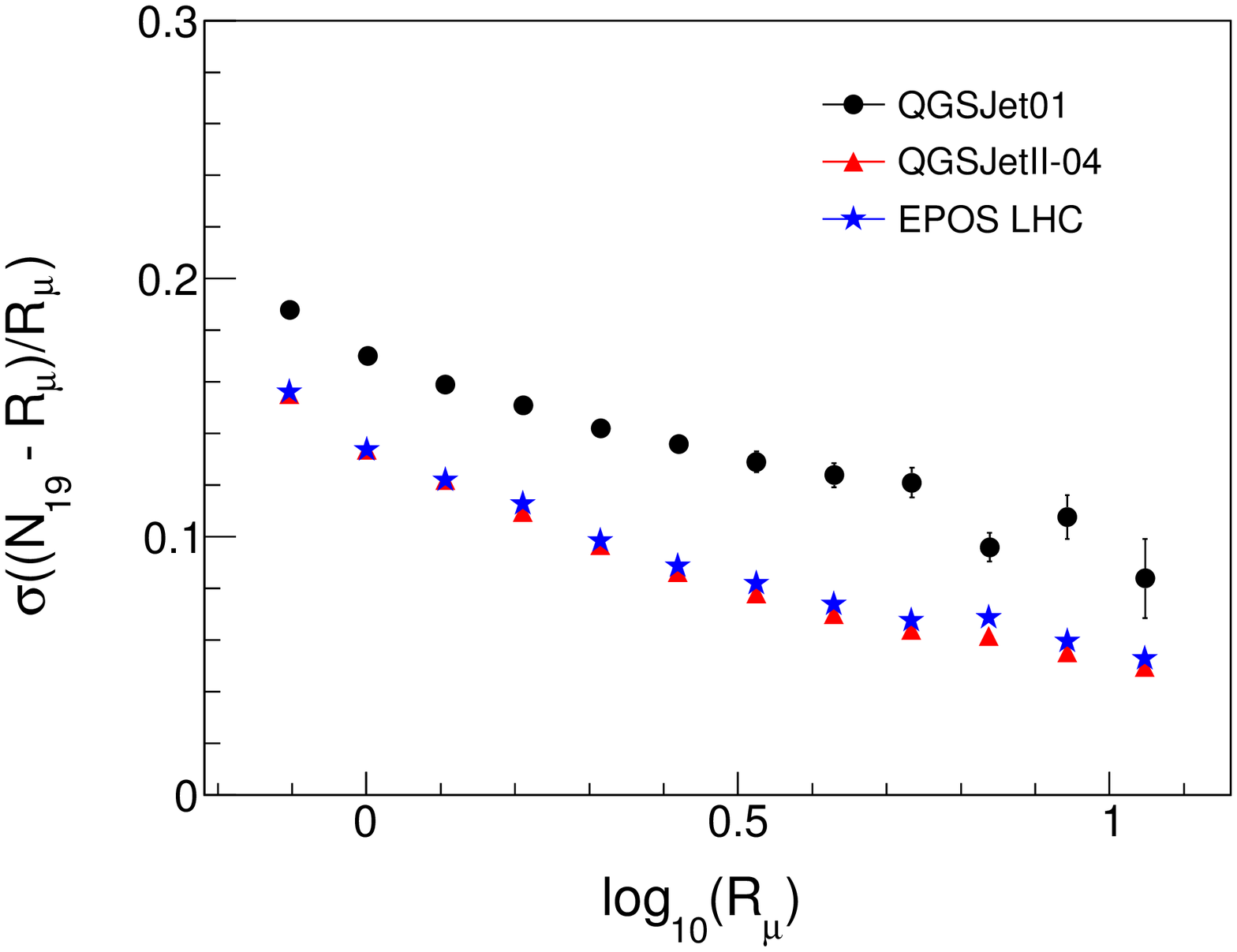}
\caption{Mean (left) and standard deviation values (right) of the relative
  difference between $N_{19}$ and $R_\mu$ (see text), as a function of $R_\mu$
  for simulated proton events.}
\label{DeltaN19}
\end{figure}

The difference between $N_{19}$ and $R_\mu$, and its standard deviation, are
illustrated in the left and right panels of figure~\ref{DeltaN19},
respectively, for protons in a variety of models. The relative difference is
on average less than 6\%. This implies that $N_{19}$ is a good estimator of
the number of muons in the shower. The relative difference observed can be
interpreted as a bias in the measurement of the number of muons. It can be
corrected for if $N_{19}$ is to be used as a measurement of the number of
muons in the shower~\cite{InesICRC,MuonPaper}.  However, such a bias has no
relevance for the spectrum calculation, since the energy is finally obtained
with the calibration procedure using events with their energy determined by
the FD.  The standard deviation decreases as the shower size increases, since
for larger showers there are more stations entering the fit. Relative standard
deviation ranges from 16 to 19\% for the smaller shower sizes considered
($R_\mu\sim0.7$) and drops to 5 to 9\% for the largest events with
$R_{\mu}>10$.

The presented standard deviation is an upper limit of the uncertainty of the
fit to the muon distribution due to the shower-to-shower fluctuations. The
shower-to-shower fluctuations imply variations in both the total number of
muons and the shape of the muon distribution of a given shower with fixed
arrival direction, energy and primary mass. If the showers only fluctuated by
changing the scale of the muon distribution, the fitted value of $N_{19}$
would directly reflect the changes of normalization of the shower-to-shower
fluctuations.\footnote{The total number of muons fluctuates in a broad range
  between 3 and 28\%, depending primarily on composition and to a lesser
  extent on hadronic model assumptions~\cite{SizeFluc,DembinskiThesis}.  The
  minimum values are obtained with pure iron, and the maximum with a
  proton-iron mixture.} But the fluctuations in the shape of the lateral
profile are not taken into account in the fit which uses the average profile
of the muon distribution. As a result, an unknown part of the shower-to-shower
fluctuations propagates into the standard deviation shown in the right panel
of figure~\ref{DeltaN19}. This is expected to be relatively small since the
standard deviation at the larger shower sizes is well below the value of the
fluctuations in $N_\mu$ for protons, which is about 20\%.

In addition, the shown standard deviation includes the angular uncertainty
propagated into $N_{19}$, nevertheless it has only a small
effect.\footnote{This is since the angular accuracy is quite good.  A
  systematic uncertainty of $0.4^\circ$ in the zenith angle reconstruction
  corresponds to an uncertainty in $N_{19}$ of 2\% (3\%) for a shower with the
  zenith angle of $60^\circ$ ($80^\circ$).} This is consistent with the
uncertainties in $N_{19}$ obtained in the fitting procedure and shown in the
right panel of figure~\ref{Sbias}.

The standard deviation is slightly lower (by ${\sim}3\%$) for showers
simulated with \textsc{QGSJetII-04} and \textsc{Epos\,LHC} than for those
simulated with \textsc{QGSJet01}. This is possibly due to shower fluctuations
and to the fact that the reference muon distributions were simulated with
\textsc{QGSJetII-03}.

\subsubsection{Systematic uncertainties}
\label{sec_sys}

The EM correction to the detector signal (section~\ref{sec_emcomponent})
depends on the primary energy, on the composition and on the hadronic
interaction model, and becomes largest in the reconstruction of shower size at
zenith angles close to $60^\circ$.  The actual value obtained for $N_{19}$
depends upon the particular choice of energy, composition and hadronic model
made for the reference EM correction, which correspond to proton simulations
at $10^{19}$\,eV with \textsc{QGSJet01} in this work. The systematic effect in
$N_{19}$ associated with the unknown composition and the hadronic model was
evaluated by computing alternative EM corrections (as defined in
eq.~\eqref{EMratio}) with simulated showers using different combinations of
energy ($10^{18}$ and $10^{20}$\,eV), composition (proton and iron) and
interaction models (\textsc{QGSJet01} and \textsc{Sibyll}). A sample of data
was reconstructed using these variations of the EM correction, leading to
corresponding new values of the shower size, $N_{19}^\text{corr}$.  The
relative differences in shower size with respect to the reference provide
estimates of the systematic uncertainties.

\begin{figure}[tbp]
\centering
\includegraphics[width=0.48\textwidth]{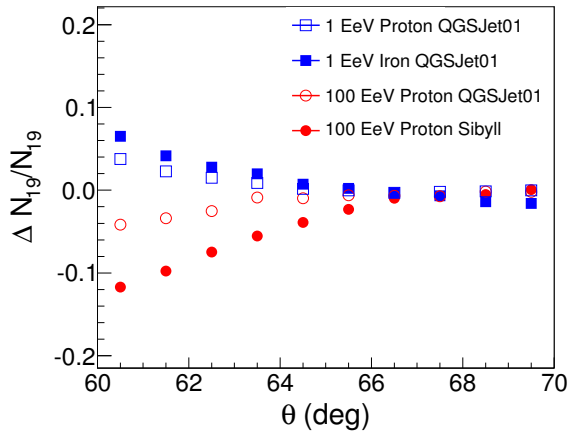}
\caption{Systematic uncertainties in shower size of a data sample obtained by
  taking the difference between reconstructed values of $N_{19}$, using the
  default correction (10\,EeV proton showers with \textsc{QGSJet01}), and
  alternative corrections derived from other simulations (as labeled). The
  energy dependence of the systematic effect is illustrated for 1\,EeV (open
  squares) and 100\,EeV (open circles) proton showers. Other scenarios are
  also shown for comparison (as labeled).}
\label{fig1} 
\end{figure}

In figure~\ref{fig1} the relative differences for several scenarios are shown
as a function of the shower zenith angle. The largest uncertainty in $N_{19}$
corresponds to the case of the correction computed with 100\,EeV proton
showers simulated with \textsc{Sibyll}. It decreases from 12\% at
$\theta\sim60^\circ$ to less than 3\% in absolute value at ${>}65^\circ$. This
is a conservative estimate since it corresponds to an extreme situation in
which the majority of data having energies of a few EeV are reconstructed with
the electromagnetic correction corresponding to a 100\,EeV shower.

There can be other possible systematic effects on the shower size which depend
on zenith angle and are not absorbed in the calibration procedure. Different
hadronic interaction models and primary compositions may have different muon
attenuations that is manifested in the zenith angle dependence of the muon
distributions. These have an intrinsic dependence on the zenith angle which
could in principle differ from that of the data. In addition, there are
implicit uncertainties which can have zenith angle, dependence due to the
accuracy of the models used for the muon distributions, the detector response
and, most importantly, the electromagnetic correction.

The extent of these systematic uncertainties can be tested by exploring the
zenith angle distribution of events above given thresholds of shower size. If
the arrival directions are isotropically distributed and the detector has an
efficiency that is independent of zenith angle $\theta$, the events should
have a flat distribution in $\sin^2\theta$. Possible systematic effects
associated with the zenith angle will appear as deviations from a uniform
distribution. This test is quite sensitive due to the steeply falling
spectrum. For a spectral index $\gamma$ it can be shown that a relative
systematic shift of shower size $\Delta_{N_{19}}^\text{sys}/N_{19}$ in a given
zenith angle bin should lead to a relative increase in the corresponding bin
of ${\sim}(\gamma-1)\,\Delta_{N_{19}}^\text{sys}/N_{19}$.

\begin{figure}[tbp]
\centering
\includegraphics[width=0.48\textwidth]{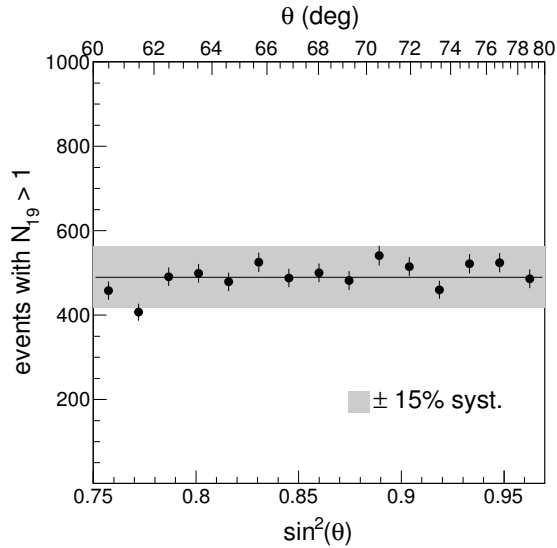}
\caption{Distribution of data events with $N_{19}>1$ in $\sin^2\theta$ bins. A
  shaded band of 15\% is shown to account for systematic uncertainties
  associated with the zenith angle in the models (see text).}
\label{Sin2ThetaPlot} 
\end{figure}

Figure~\ref{Sin2ThetaPlot} displays the distribution of events with $N_{19}>1$
in bins of equal geometrical exposure. The array is fully efficient for
showers of this size.  The plot indicates that systematic deviations are
within a 15\% band with a standard deviation of ${\sim}8\%$. Assuming a
spectral index $\gamma\simeq2.69\pm0.02$ as obtained with events below
$60^\circ$, and for energies above 4\,EeV~\cite{SpectrumPRL} this systematic
uncertainty corresponds to a potential shift in $N_{19}$ of less than
$\pm9\%$, well within the combined estimated uncertainties of the muon
distributions, detector response and electromagnetic corrections.

\section{Energy calibration and resolution}
\label{s:energy_calibration}

It is possible to relate $N_{19}$ to the shower energy using simulations, but
due to the lack of knowledge of composition and hadronic models the
corresponding systematic uncertainties become quite large (see
section~\ref{s:number_density}). Alternatively, the correlation between
$N_{19}$ and shower energy reconstructed with the FD can be obtained from a
subset of ``golden hybrid'' events (events for which both FD and SD
reconstructions are possible), similarly to what has been used to calibrate
events below $60^\circ$~\cite{SpectrumPRL}. The energy scale inferred from
this subset is applied to all the inclined showers recorded with the SD.

The golden hybrid events with zenith angle greater than $60^\circ$ are
required to pass the inclined T4 and T5 conditions and, in addition, to
satisfy a set of FD quality cuts specifically designed to ensure an accurate
reconstruction of the arrival direction and of the longitudinal profile. The
cuts are adapted versions of those used in the calibration of events with
$\theta<60^\circ$~\cite{Pesce}. The station closest to the shower core which
is used for the geometrical reconstruction must be at a distance less than
750\,m. For a precise estimate of the energy, we require an adequate
monitoring of the atmospheric conditions (vertical aerosol optical depth up to
0.1; cloud coverage less than 25\% in the FD field of view, distance of the
cloud layer to the measured profile greater than 50\,g/cm$^2$, and thickness
of the cloud layer less than 100\,g/cm$^2$). Furthermore, we exclude a
residual contamination of shower profiles distorted by clouds and aerosols by
requiring a Gaisser-Hillas fit with a residual
$(\chi^2-n_\text{dof})/\sqrt{2n_\text{dof}}$ smaller than 3 and a negative
value of the parameter $X_\text{0}$ of the fitted Gaisser-Hillas
profile.\footnote{In the Gaisser-Hillas function, $X_\text{0}$ is a parameter
  not to be confused with the depth of the first interaction. Showers in both
  data and simulation are found to be best described by negative
  values~\cite{SongGH,HIRESGH}.} Moreover, the maximum accepted uncertainty of
$X_\text{max}$ is 150\,g/cm$^2$. In addition to the quality selection
criteria, a fiducial cut on the FD field of view (FOV) is
applied~\cite{Abraham:2010yv}, ensuring that it is large enough to observe all
plausible values of the shower maximum $X_\text{max}$. This ``fiducial FOV
cut'' includes a restriction on the minimum viewing angle of the light in the
FD telescope ($25^\circ$).  Finally, only events with FD energies greater than
$4{\times}10^{18}$\,eV are accepted to ensure a trigger probability of nearly
100\% for the SD and FD detectors. For the period from 1 January 2004 to 31
December 2012 the sample has 223 hybrid events with $\theta\geq60^\circ$.

To describe the correlation it is sufficient to perform a power-law fit to the
shower size $N_{19}$ as a function of the calorimetric hybrid energy
$E_\text{FD}$,
\begin{equation}
 N_{19}=A\,(E_\text{FD}/10^{19}\,\text{eV})^B,
\label{correlation} 
\end{equation}
which is then inverted to give the energy conversion. The slope parameter $B$
is related to $\alpha$ ($\rho_\mu\propto E^\alpha$) as discussed in
section~\ref{s:number_density}. The fit must be handled with care since it is
performed on a subset of the data used to calculate the spectrum. A bias can
be expected at low energies due to the threshold trigger effects of the SD
stations.

\begin{figure}[tbp]
\centering
\includegraphics[height=0.5\textwidth]{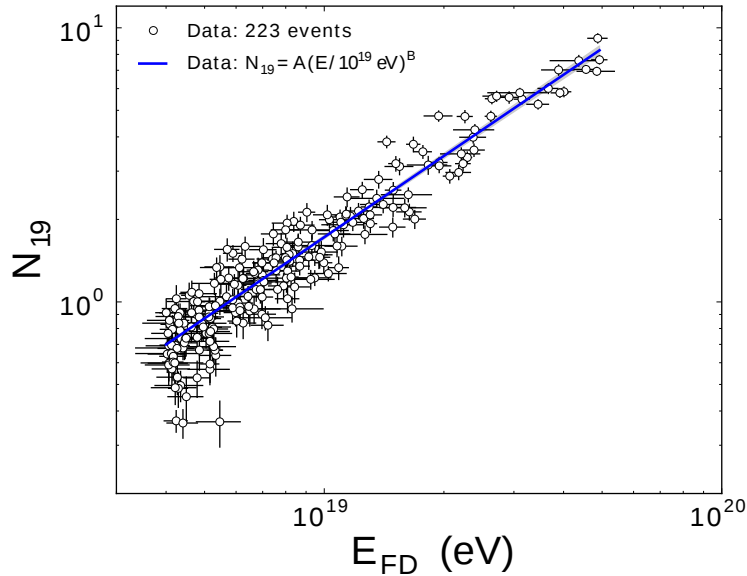}
\caption{Correlation between $N_{19}$ and $E_\text{FD}$ for golden hybrid data
  with $\theta\geq60^\circ$. The solid line is the best fit of the power-law
  $N_{19}=A\,(E_\text{FD}/10^{19}\,\text{eV})^B$ to the data (see text).}
\label{calibration} 
\end{figure}

The fit is based on a tailored maximum-likelihood method~\cite{ICRC11Hans}
that takes into account the effect of the cut on energies greater than
$4{\times}10^{18}$\,eV, which would otherwise distort a standard least-squares
fit. The ability to include both the uncertainties of the reconstructions of
$N_{19}$ (see figure~\ref{Sbias} in section~\ref{ss:recprocedure}) and
$E_\text{FD}$, without relying on approximations, is another advantage of this
approach. We achieve this by regarding the data as drawn from a
two-dimensional probability density function $f(E_\text{FD},N_{19})$ that
describes the random fluctuations away from the ideal curve given by
eq.~\eqref{correlation}. The PDF is centered around this curve and its width
is given by the uncertainties of the reconstructed values of $N_{19}$ and
$E_\text{FD}$. Shower-to-shower fluctuations of $N_{19}$ also contribute to
the spread and are taken into account. Their relative size is assumed to be
constant in the energy range of interest and fitted to the data as an extra
parameter. Since the data are described by a two-dimensional PDF, we are able
to model the effect of an energy cut by setting the probability to observe
events below the cut to zero. The effect of event migration in and out of the
accepted region is also treated correctly. The method was validated with
extensive Monte-Carlo studies and was found to fit the calibration curve
without bias. The details will be presented in a dedicated article.

The result of this fit is illustrated in figure~\ref{calibration} and the best
resulting parameters are $A=1.723\pm0.023$ and $B=0.984\pm0.020$.  The
calibration accuracy at the highest energies is limited by the number of
events (the most energetic is at ${\sim}5{\times}10^{19}$\,eV). An alternative
fitting method based on least-squares minimization was used to calibrate the
selected hybrid set as a cross-check. In this method the trigger bias is
removed with an elliptical cut in the shower-size-energy plane to ensure that
trigger effects can be ignored. The parameters obtained from the best fit are
compatible with the maximum-likelihood fit.

\begin{figure}[tbp]
\centering
\includegraphics[width=0.5\textwidth]{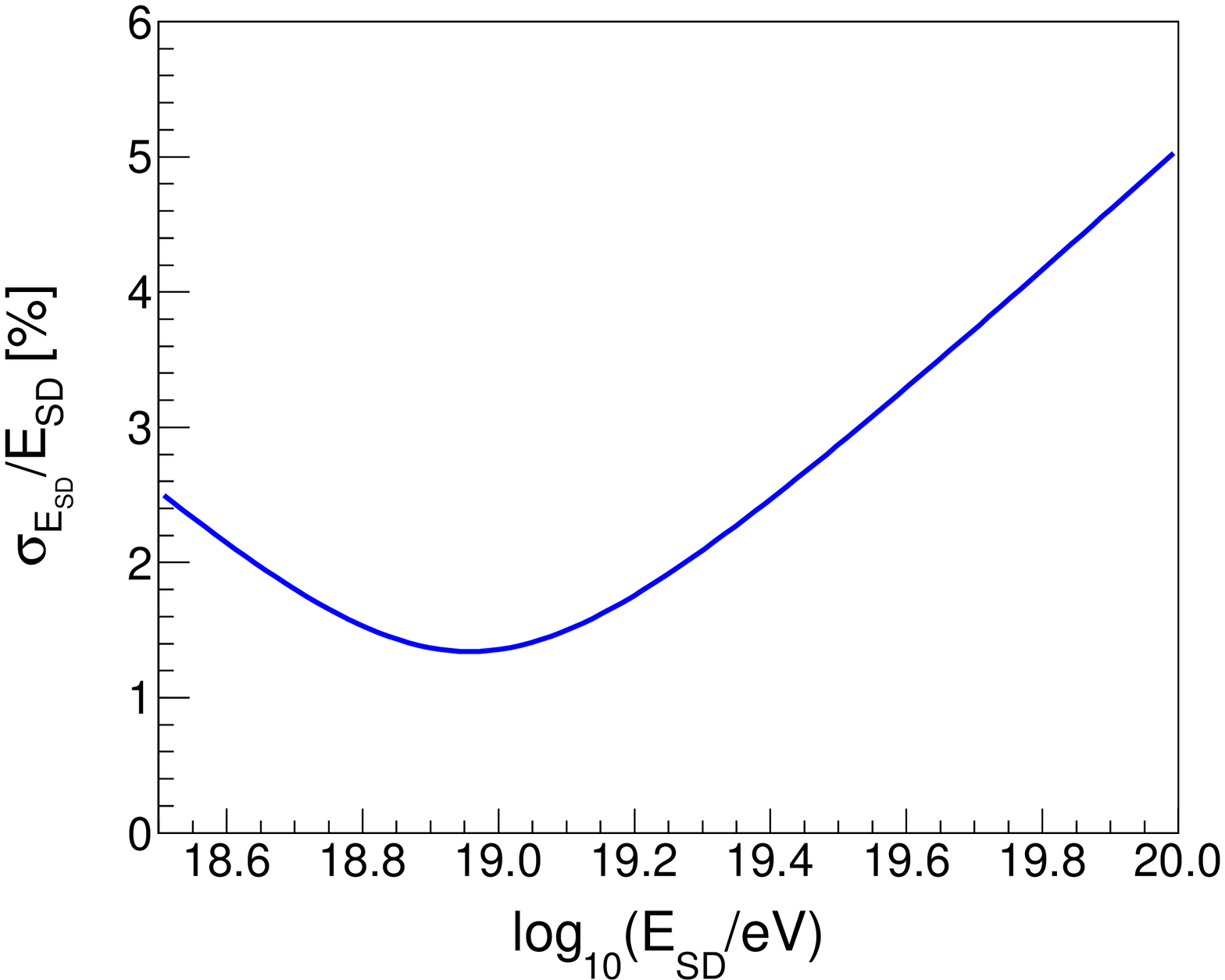}
\hfill
\includegraphics[width=0.44\textwidth,clip=]{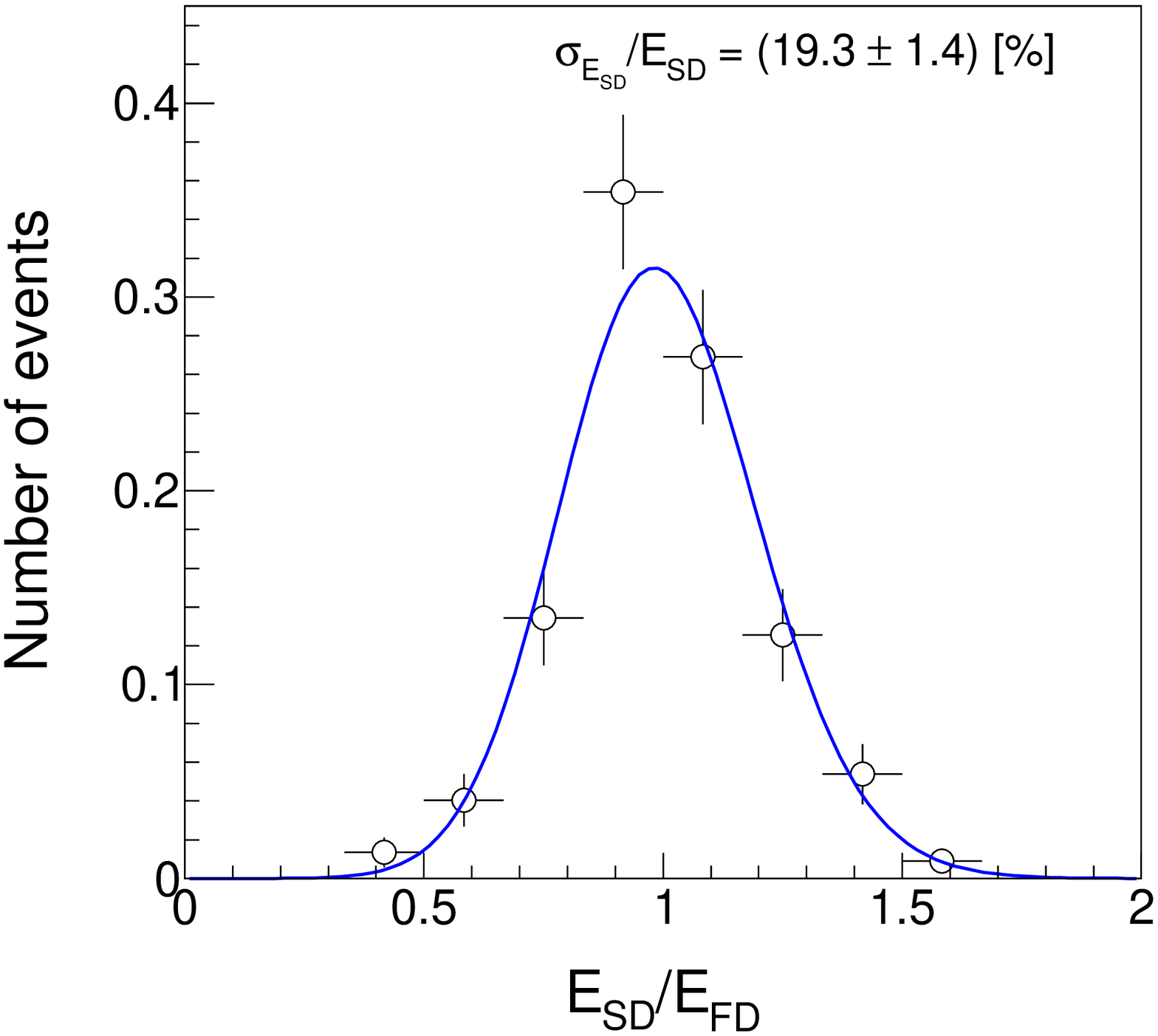}
\caption{\emph{Left:} relative systematic uncertainty in $E_\text{SD}$ as a
  function of shower energy as obtained by propagating the uncertainties from
  the correlation fit. \emph{Right:} energy resolution inferred from the
  distribution of the ratio between calibrated SD energy $E_\text{SD}$ and FD
  energy $E_\text{FD}$ for all the golden hybrid data used in the
  calibration.}
\label{N19Pull} 
\end{figure}

Eq.~\eqref{correlation} can be used to convert $N_{19}$ to shower energy
$E_\text{SD}$ for all the inclined events recorded with the SD. The
statistical uncertainties of the calibration constants $A$ and $B$ were
converted to uncertainties in $E_\text{SD}$ in the left panel of
figure~\ref{N19Pull}. The latter range between 1.5\% at $10^{19}$\,eV and 5\%
at $10^{20}$\,eV.

To be detected with the FD the showers must have a depth of maximum well
within the field of view, and to be detected with the SD the shower core must
fall inside the array. The two simultaneous conditions suppress events with
large zenith angles due to the geometrical layout of the
Observatory~\cite{guerard}. The systematic uncertainty in $E_\text{SD}$,
accounting for the different angular distributions of the golden hybrid events
and the full inclined data set used to calculate the spectrum, is on average
${\sim}2\%$.

In addition, there is an overall systematic uncertainty of 14\% from the FD
energy measurement which relies on the knowledge of the fluorescence yield
used in the energy determination of the FD events (3.6\%), atmospheric
conditions (3.4 to 6.2\%), absolute detector calibration (9.9\%), invisible
energy (3 to 1.5\%), stability of the energy scale (5\%), and shower
reconstruction (6.5 to 5.6\%)~\cite{AbsFDunc}.

The total systematic uncertainty in $E_\text{SD}$ ranges between 14\% at
$10^{19}$\,eV and 17\% at $10^{20}$\,eV, and is obtained by adding the
uncertainties described above in quadrature.

The right panel of figure~\ref{N19Pull} illustrates the ratio of the inferred
SD energy $E_\text{SD}$ (after the calibration procedure) and the
reconstructed FD-Hybrid energy $E_\text{FD}$.  The resolution in the SD energy
can be inferred from this ratio distribution~\cite{Geary}, by fixing the FD
energy resolution to 7.6\%~\cite{AbsFDunc}. The resulting average SD energy
resolution is $(19.3\pm1.4)\%$ for the selected golden hybrid set. This SD
energy resolution is attributed to the combined effect of shower-to-shower
fluctuations and the reconstruction uncertainty of the fitting procedure (see
figure~\ref{Sbias} in section~\ref{ss:recprocedure}).

The fact that the value of the calibration parameter $A$ is not exactly equal
to one reflects the fact that the signals detected with the SD are larger than
expected from the reference distribution using protons. This is similar to the
situation for events at zenith angles less than $60^\circ$, where the measured
signal exceeds that expected for proton showers at the same energy, but the
energy of the muons involved in inclined showers is significantly
higher~\cite{model}. The obtained value of $A$ implies that $N_{19}=1$
corresponds to an FD energy of 5.75\,EeV, about 42\% less than the energy of
10\,EeV used for the proton reference distributions.  Equivalently, it implies
that the number of muons in a 10\,EeV shower is larger than the average muon
content of the reference proton showers simulated with
\textsc{QGSJetII-03}~\cite{ICRC11GRF}. This result has important implications
concerning composition and hadronic models, which will be addressed in a
separate article.

\section{Summary and conclusions}
\label{s:conclusions}

We have developed a procedure to reconstruct inclined showers detected with
the SD of the Pierre Auger Observatory. After reconstructing the arrival
direction in a standard way using the start times of the recorded signals in
the triggered stations, the signals of each event are fitted to the expected
shape of the muon distribution at the ground.  This requires modeling of the
corresponding muon patterns, the response of the detectors to the passage of
particles, and the amount of electromagnetic component of inclined showers,
all of which were obtained with the aid of comprehensive simulations.  The
reference patterns for the muon distribution for each arrival direction are
based on proton simulations with \textsc{QGSJetII.03} at a fixed energy of
10\,EeV.  A size parameter $N_{19}$ has been introduced to scale the SD
measurements to the reference distributions. $N_{19}$ is used as an energy
estimator, calibrating it with the calorimetric energy measured by the FD,
using a sub-sample of quality hybrid events which can be reconstructed
independently using both the FD and SD techniques.  In this process the
simulated proton showers used to obtain the reference distributions of muons
at the ground are shown to have a large deficit of muons with respect to the
measured data.

The performance of the reconstruction was studied for events in the zenith
angle range between $60^\circ$ and $80^\circ$, and with $N_{19}$ exceeding
0.7, to ensure nearly full efficiency for the trigger, selection and
reconstruction of events that fall in active regions of the array. The angular
resolution was shown to be better than $0.5^\circ$.  The SD energy resolution
was obtained, comparing the reconstructed SD and FD energies for the subset of
events used in the energy calibration. The average SD energy resolution was
shown to be about 19.3\% attributed both to the reconstruction procedure and
to the shower-to-shower fluctuations. Systematic uncertainties due to the
calibration procedure were estimated to be energy dependent, and to fall in
the range from 1.5 to 5\%. In addition, there is an overall systematic
uncertainty of 14\% associated with the FD energy assignment.  The procedure
to analyze inclined events has a resolution in size and angular accuracy which
is comparable to that obtained for events with $\theta<60^\circ$. The analysis
of these events opens the possibility to measure the cosmic ray spectrum in an
independent way, to explore primary composition, to study the fidelity of
current hadronic models extrapolated to the high energies relevant to these
events, and to analyze arrival directions from parts of the sky that are not
accessible using events with zenith angles less than $60^\circ$.  An
independent measurement of the energy spectrum of ultra-high energy cosmic rays using
very inclined showers is to be presented in a forthcoming publication.

\acknowledgments

The successful installation, commissioning, and operation of the Pierre Auger
Observatory would not have been possible without the strong commitment and
effort from the technical and administrative staff in Malarg\"{u}e.

We are very grateful to the following agencies and organizations for financial
support: Comisi\'{o}n Nacional de Energ\'{\i}a At\'{o}mica, Fundaci\'{o}n
Antorchas, Gobierno De La Provincia de Mendoza, Municipalidad de Malarg\"{u}e,
NDM Holdings and Valle Las Le\~{n}as, in gratitude for their continuing
cooperation over land access, Argentina; the Australian Research Council;
Conselho Nacional de Desenvolvimento Cient\'{\i}fico e Tecnol\'{o}gico (CNPq),
Financiadora de Estudos e Projetos (FINEP), Funda\c{c}\~{a}o de Amparo \`{a}
Pesquisa do Estado de Rio de Janeiro (FAPERJ), S\~{a}o Paulo Research
Foundation (FAPESP) Grants \# 2010/07359-6, \# 1999/05404-3, Minist\'{e}rio de
Ci\^{e}ncia e Tecnologia (MCT), Brazil; MSMT-CR LG13007, 7AMB14AR005,
CZ.1.05/2.1.00/03.0058 and the Czech Science Foundation grant 14-17501S, Czech
Republic; Centre de Calcul IN2P3/CNRS, Centre National de la Recherche
Scientifique (CNRS), Conseil R\'{e}gional Ile-de-France, D\'{e}partement
Physique Nucl\'{e}aire et Corpusculaire (PNC-IN2P3/CNRS), D\'{e}partement
Sciences de l'Univers (SDU-INSU/CNRS), France; Bundesministerium f\"{u}r
Bildung und Forschung (BMBF), Deutsche Forschungsgemeinschaft (DFG),
Finanzministerium Baden-W\"{u}rttemberg, Helmholtz-Gemeinschaft Deutscher
Forschungszentren (HGF), Ministerium f\"{u}r Wissenschaft und Forschung,
Nordrhein Westfalen, Ministerium f\"{u}r Wissenschaft, Forschung und Kunst,
Baden-W\"{u}rttemberg, Germany; Istituto Nazionale di Fisica Nucleare (INFN),
Ministero dell'Istruzione, dell'Universit\`{a} e della Ricerca (MIUR), Gran
Sasso Center for Astroparticle Physics (CFA), CETEMPS Center of Excellence,
Italy; Consejo Nacional de Ciencia y Tecnolog\'{\i}a (CONACYT), Mexico;
Ministerie van Onderwijs, Cultuur en Wetenschap, Nederlandse Organisatie voor
Wetenschappelijk Onderzoek (NWO), Stichting voor Fundamenteel Onderzoek der
Materie (FOM), Netherlands; National Centre for Research and Development,
Grant Nos.ERA-NET-ASPERA/01/11 and ERA-NET-ASPERA/02/11, National Science
Centre, Grant Nos. 2013/08/M/ST9/00322 and 2013/08/M/ST9/00728, Poland;
Portuguese national funds and FEDER funds within COMPETE - Programa
Operacional Factores de Competitividade through Funda\c{c}\~{a}o para a
Ci\^{e}ncia e a Tecnologia, Portugal; Romanian Authority for Scientific
Research ANCS, CNDI-UEFISCDI partnership projects nr.20/2012 and nr.194/2012,
project nr.1/ASPERA2/2012 ERA-NET, PN-II-RU-PD-2011-3-0145-17, and
PN-II-RU-PD-2011-3-0062, the Minister of National Education, Programme for
research - Space Technology and Advanced Research - STAR, project number
83/2013, Romania; Slovenian Research Agency, Slovenia; Comunidad de Madrid,
FEDER funds, Ministerio de Educaci\'{o}n y Ciencia, Xunta de Galicia, Spain;
The Leverhulme Foundation, Science and Technology Facilities Council, United
Kingdom; Department of Energy, Contract No. DE-AC02-07CH11359,
DE-FR02-04ER41300, and DE-FG02-99ER41107, National Science Foundation, Grant
No. 0450696, The Grainger Foundation, USA; NAFOSTED, Vietnam; Marie
Curie-IRSES/EPLANET, European Particle Physics Latin American Network,
European Union 7th Framework Program, Grant No. PIRSES-2009-GA-246806; and
UNESCO.

\end{document}